\documentclass[journal,12pt,onecolumn,draftclsnofoot,]{IEEEtran}
\usepackage{cite}
\usepackage{graphicx,color,epsfig,rotating}
\usepackage{amsfonts,amsmath,amssymb,bbm}
\usepackage{algorithm, setspace}
\usepackage{algpseudocode}
\usepackage{subfigure}
\usepackage{amsmath}
\usepackage{cite}
\usepackage{mdwtab}
\usepackage{placeins}
\usepackage{psfrag, graphicx}
\usepackage[latin1]{inputenc}
\usepackage{amssymb}
\usepackage{multirow}
\usepackage{stfloats}
\usepackage{tabularx} 
\usepackage{booktabs} 
\usepackage{url}
\usepackage{bm}
\usepackage{float}

\newcommand{\cq}{{\mathcal{Q}}}
\newcommand{\ttt}{{t_T}}
\newcommand{\ttr}{{t_R}}
\newcommand{\mt}{{M_T}}
\newcommand{\mr}{{M_R}}
\newcommand{\kt}{{K_T}}
\newcommand{\kr}{{K_R}}
\newcommand{\dt}{{D_T}}
\newcommand{\dr}{{D_R}}
\newcommand{\cw}{{\mathcal{W}}}
\newcommand{\tron}{{\frac{\mt\kt +\mr\kr}{N}}}

\newcommand{\fr}{{\binom{\dr -2}{\delta -1}\binom{\dr -1}{\delta}^{\ttr -1}\frac{(\delta !)^{\ttr}}{\delta}(\ttr -1)!    }}  

\newfont{\bbb}{msbm10 scaled 700}

\newfont{\bb}{msbm10 scaled 1100}

\newcommand{\Wc}{{\cal W}}

\newcommand{\msf}{{\sf m}}

\newcommand{\be}{\begin{equation}}
\newcommand{\ee}{\end{equation}}
\newcommand{\bea}{\begin{eqnarray}}
\newcommand{\eea}{\end{eqnarray}}

\let\tbf\textbf
\let\tit\textit
\let\mc\mathcal

\let\mbb\mathbb

\let\msf\mathsf

\newtheorem{defn}{Definition}
\newtheorem{example}{Example}
\newtheorem{theorem}{Theorem}
\newtheorem{lemma}{Lemma}
\makeatletter
\newcommand{\subalign}[1]{%
  \vcenter{%
    \Let@ \restore@math@cr \default@tag
    \baselineskip\fontdimen10 \scriptfont\tw@
    \advance\baselineskip\fontdimen12 \scriptfont\tw@
    \lineskip\thr@@\fontdimen8 \scriptfont\thr@@
    \lineskiplimit\lineskip
    \ialign{\hfil$\m@th\scriptstyle##$&$\m@th\scriptstyle{}##$\crcr
      #1\crcr
    }%
  }
}
\makeatother

\usepackage{changes}
\definechangesauthor[color=blue,name={Mingyue Ji}]{JMY}

\ifodd 1

\else

\fi

\ifodd 1

\else

\fi

\begin{document}
\title{Cache-aided Interference Management using Hypercube Combinatorial
Design with Reduced Subpacketizations and Order Optimal Sum-Degrees of Freedom}

\author{Xiang Zhang,~\IEEEmembership{Student Member,~IEEE}, Nicholas Woolsey,~\IEEEmembership{Student Member,~IEEE},\\ and Mingyue Ji,~\IEEEmembership{Member,~IEEE}
\thanks{This manuscript was partially presented in the conference papers \cite{zhang2019icc}.}
\thanks{The authors are with the Department of Electrical Engineering,
University of Utah, Salt Lake City, UT 84112, USA. (e-mail: xiang.zhang@utah.edu, nicholas.woolsey@utah.edu, and mingyue.ji@utah.edu)}
}


\maketitle

\vspace{-1cm}
\begin{abstract}
We consider a cache-aided interference network which consists of a library of $N$ files, $K_T$ transmitters and $K_R$ receivers (users), each equipped with a local cache of size $M_T$ and $M_R$ files respectively, and connected via a discrete-time additive white Gaussian noise (AWGN) channel. Each receiver requests an arbitrary file from the library. The objective is to design a cache placement without knowing the receivers' requests and a communication scheme such that the sum Degrees of Freedom (sum-DoF) of the delivery is maximized. This network model with one-shot transmission was firstly investigated by Naderializadeh {\em et al.}, who proposed a 
scheme that achieves a one-shot sum-DoF of $\min\{\frac{{M_TK_T+K_RM_R}}{{N}}, K_R\}$, which is optimal within a constant of $2$. One of the biggest limitations of this scheme is the requirement of high subpacketization level. This paper attempts to design new algorithms to reduce the file subpacketization in such a network without hurting the sum-DoF.
In particular, we propose a new approach for both prefetching and linearly coded delivery based on a combinatorial design called {\em hypercube}. The proposed approach reduces the subpacketization exponentially in terms of $K_R M/N$ ({$M=M_T$ or $M_R$ represents the transmitter/receiver cache size}) and achieves the identical one-shot sum DoF when $\frac{M_TK_T+K_RM_R}{N} \leq K_R$.   
\end{abstract}



\section{Introduction}
\label{section:intro}
Wireless traffic has grown dramatically in recent years due to the increasing mobile data demand, mainly due to video delivery services \cite{cisco2016global}. One promising approach to 
handle this traffic bottleneck is to exploit local cache memories at end user devices or network edge nodes (e.g., small cell base stations) to pre-store part of the contents (e.g, movies) which might be requested in the near future. With the help of these cache nodes, the system can serve users with a much higher rate and lower latency \cite{maddah2014fundamental,wan2016caching,ji2016fundamental,
maddahali2015interference,Naderializadeh2017interference,Hachem2018interference,
bidokhti2017gaussian, Sengupta2017fog,shariatpanahi2017physical,xu2017caching,lampiris2018adding,xu2018fog}. 
Among all schemes based on caching approaches, coded caching, introduced in \cite{maddah2014fundamental}, has attracted significant attentions. In particular, Maddah-Ali and Niesen considered a {\em shared-link network} and studied the problem of minimizing the worst-case traffic load or {\em transmission rate}. It was shown that prefetching packets of the library files in a uniform manner during the placement phase, and employing coded scheme based on linear index code during the delivery phase, is sufficient to provide optimal rate under uncoded cache placement \cite{wan2016caching,yu2018uncoded,wan2020index}. Later, the idea of coded caching was extended to many other network topologies including Device-to-Device (D2D) caching networks \cite{ji2016fundamental}, 
 multi-server caching networks \cite{Shariatpanahi2016multiserver} and combination caching networks \cite{ji2015fundamental,ji2015combination,wan2017survey,8374866,wan2020combination}, where the channels between the transmitters and receivers are either wireline channel or noiseless broadcast channel. 

The concept of coded caching was also extended to the wireless channels with the 
consideration of interference
\cite{maddahali2015interference,bidokhti2017gaussian,Naderializadeh2017interference,Hachem2018interference,xu2017caching,Sengupta2017fog,shariatpanahi2017physical,lampiris2018adding}. 
For example, in \cite{maddahali2015interference}, the authors considered a three-user interference channel where only transmitters are equipped with cache memories (no cache memories at the receivers) and showed that via a specific cache prefetching strategy, an efficient delivery scheme can be designed by exploiting the gains based on interference cancellation and interference alignment. {In  \cite{bidokhti2017gaussian}, the additive Gaussian channel in a broadcast setting with cache-aided receivers was studied.} Later, the study was extended to the case where both transmitters and receivers are equipped with cache memories \cite{Naderializadeh2017interference,Hachem2018interference,xu2017caching,lampiris2018adding}. Moreover, cache-aided fog radio access network was also investigated in \cite{Sengupta2017fog,xu2018fog}. 

As shown in the above works, in most of the network models, the remarkable multiplicative gain of coded caching in terms of network aggregate cache memory has been established in the asymptotic regime when the number of packets per file, denoted by $F$, scales to infinity. It has been shown that in most of the cases, to achieve the desired caching gain, $F$ has to increase exponentially as a function of the number of nodes in the network. The finite length analysis of coded caching for the shared-link network was initiated in \cite{Shanmugam2016} in which the authors proposed to encode the data only across a small subset of the total $K$ users in the system to obtain reduced subpacketization level at the cost of a reduced 
 coded caching gain. Significant efforts have been made to reduce the subpacketization levels in shared-link caching networks such as placement delivery array (PDA) \cite{yan2017pda}, resolvable design \cite{tang2018subpacketization} and hypergraph-based design \cite{shangguan2018hpyergraph}.

{The finite length analysis of coded caching in other network topologies other than shared-link and MIMO broadcast channel is very limited. {In \cite{lampiris2018adding}, the authors considered a MISO broadcast channel with $L$ transmit antennas and showed that a reduced subpacketization can be achieved}. In addition, the scheme was applied to the cache-aided interference setting \cite{Naderializadeh2017interference} where asymmetric cache placements are used in the transmitter and receiver sides. However, due to receiver grouping, the scheme in \cite{lampiris2018adding} can not achieve the optimal sum-DoF of $K_RM_R/N+L$ when either $K_R/L$ or $K_RM_R/NL$ is not an integer, limiting it applicability to networks with large number of transmit antennas. When applied to the interference setting, this scheme can only achieve the order optimal sum-DoF when the transmit caching parameter $K_TM_T/N$ is smaller than the receiver caching parameter $K_RM_R/N$. A later work \cite{salehi2020low}  
studied the complexity issues of coded caching in MISO systems with large number of transmit antennas. A cyclic cache placement was designed using PDA \cite{yan2017pda} and a quadratic (w.r.t. number of receivers) subpacketization can be achieved. However, the proposed scheme can only achieve the optimal DoF when the number of transmit antennas is larger than the receiver caching parameter, i.e., $L\ge K_RM_R/N$, which puts a  limitation to its applicability to interference network settings.} 
In \cite{woolsey2020d2d}, we considered a D2D caching network over noiseless broadcast channel model and introduced a combinatorial design called {\em hypercube}, and the corresponding placement and coded delivery schemes with a substantially lower subpacketization level while still achieving order optimal throughput.  

In this paper, we consider the general wireless interference network with cache memories equipped at both the transmitter and receiver sides. In particular, we consider a wireless interference network with $K_T$ transmitters and $K_R$ receivers, each equipped with a local cache memory of size $M_T$ and $M_R$ files, from a library of $N$ files. We restrict the communication scheme to one-shot linear schemes due to its practicality. This network model was first considered by Naderializadeh, Maddah-Ali and Avestimehr (NMA) in \cite{Naderializadeh2017interference}.  
Interestingly, in this work, we will show that our previously introduced hypercube based combinatorial approach, which was designed for D2D caching networks with noiseless broadcast channels, can be extended to cache-aided interference networks in a non-straightforward way such that the subpacketizations can be significantly reduced. 
{The main challenge of adapting the hypercube cache placement to the $K_T\times K_R$ interference network lies in the design of the delivery phase. The difference of this work from \cite{woolsey2020d2d} is that, since only the shared-link was considered in \cite{woolsey2020d2d}, only the cache-induced interference cancellation opportunities (coded caching gain) are available. However, in our considered setting, through transmitter cooperation, zero forcing is also available and therefore the delivery scheme of \cite{woolsey2020d2d} can not be used here and new delivery schemes have to be designed to maximally exploit both coded caching and zero forcing gain.}

Our main contribution in this paper is two-fold. First, based on the hypercube cache placement introduced in \cite{woolsey2020d2d}, we designed a cache placement scheme at both transmitters and receivers, and proposed a linear one-shot delivery scheme by exploiting zero-forcing opportunities via transmitter collaboration and cache-induced interference cancellation opportunities at receivers side. The proposed scheme achieves an order-wise subpacketization level reduction compared to that achieved in \cite{Naderializadeh2017interference}.  
Second, when $\frac{K_TM_T+K_RM_R}{N} \leq K_R$, the proposed scheme achieves a one-shot sum-DoF of $\frac{K_TM_T+K_RM_R}{N}$, which is within a factor of $2$ to the optimum as shown by \cite{Naderializadeh2017interference}.\footnote{{Note that when  $\frac{K_TM_T+K_RM_R}{N} > K_R$, using the similar argument presented in \cite{Naderializadeh2017interference}, the order optimal sum-DoF of $K_R$ is also achievable using our proposed approach. However, it is not straightforward to compare the subpacketizations. Hence, we do not consider this case in this paper.}} More importantly and surprisingly, it achieves the same sum-DoF as in \cite{Naderializadeh2017interference}. This implies that there is no loss in terms of one-shot sum-DoF by using the proposed scheme while requiring a much less file subpacketization. In the rest of the paper, we will refer the scheme in \cite{Naderializadeh2017interference} as NMA scheme. 
    
\paragraph*{Notation Convention}     
We use calligraphic symbols to denote sets and $|\cdot|$ to represent the cardinality of a set or the length of a vector. 
{$\mathbb{Z}^+$ denotes the positive integer set} and $ \mathbb{C}$ denotes the set of complex numbers. ``$a\mod b$" denotes the module operation of $a$ modulo $b$. For some $m,n\in\mathbb{Z}^+$ and $m\le n$, let $[n] \triangleq \{0, 1, \cdots, n-1\}$ and $[m:n]\triangleq \{m,m+1,\cdots,n-1,n\}$. {The standard order notations ($o(),O()$ and $\Theta()$) are used in this paper.}
    
\section{Network Model and Problem Formulation}
\label{sec: Network Model and Problem Formulation}

\subsection{General Problem Formulation}
\begin{figure}  
\centering
\includegraphics[width=0.45\textwidth]{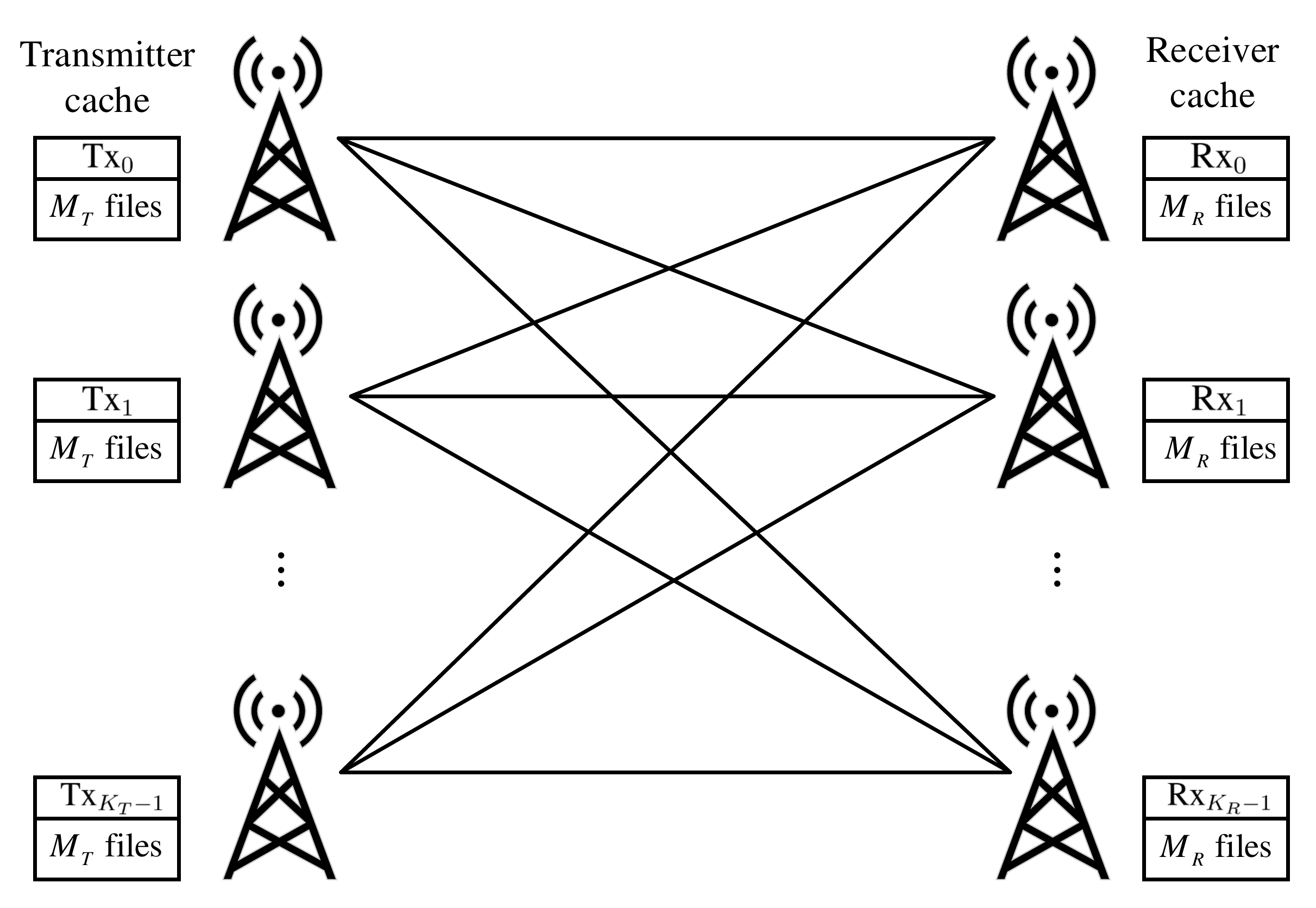}
\caption{Wireless interference network consisting of $K_T$ transmitters, each equipped with a cache of size $M_T$ files and $K_R$ receivers, each equipped with a cache of size $M_R$ files. The system also contains a library of $N$ files.}
\label{figure:1}
\end{figure}

Consider a wireless interference network, as illustrated in Fig. \ref{figure:1}, which consists of $K_T$ transmitters and $K_R$ receivers, denoted by $\{\textrm{Tx}_i: i\in[K_T]\}$ and $\{\textrm{Rx}_j: j\in[K_R]\}$, respectively. The system contains a library of $N$ files denoted by $\{\mathcal{W}_n: n\in[N]\}$, where the file $\mathcal{W}_n$ contains $F$ packets $\mathcal{W}_n\triangleq\{w_{n,p}: p\in[F]\}$ with 
size of $L$ bits each, i.e., $w_{n,p}\in \mathbb{F}_2^{L}$.\footnote{{In this paper, we let $L$ be a designed variable and equals $|\Wc_n|/F$.}} Transmitters and receivers are equipped with cache memories to store part of the file library. In particular, each transmitter and receiver are equipped with a local cache of size $M_T$ and $M_R$ files, respectively. 
The communication channel between transmitters and receivers is modeled as discrete-time additive white Gaussian noise (AWGN)  channel, which can be written as
\begin{equation}
Y_j(t)=\sum_{i=0}^{K_T-1}h_{ji}S_i(t)+N_j(t),
\end{equation}
where $t$ is the index of the time slot.\footnote{{We will ignore the index of $t$ when it does not cause confusion.}} $S_i(t)\in \mathbb{C}$ is the transmit signal of Tx$_i$ satisfying the power constraint $\mathbb{E}[|S_i(t)|^2]\leq P$. $Y_j(t)$ is the received signal of Rx$_j$ and $N_j(t)$ represents the unit-power AWGN noise 
at receiver Rx$_j$. Moreover, $h_{ji}\in\mathbb{C}$ represents the channel gain from Tx$_i$ to Rx$_j$, which is assumed to keep unchanged during the entire transmission process 
and is known to all transmitters and receivers.
The system operates in two phases: the \textit{prefetching phase} and the \textit{delivery phase} as described in \cite{Naderializadeh2017interference}. In the prefetching phase, each transmitter and receiver can store up to $M_TF$ and $M_RF$ arbitrary packets from the file library, respectively. 
This phase is done without the prior knowledge of the receivers' future requests. In the following delivery phase, 
each receiver Rx$_j$ randomly requests a 
file $\mathcal{W}_{d_j},d_j\in[N]$ from the library. These requests are represented by a \emph{demand vector} denoted as $\mathbf{d}\triangleq[d_0,d_1,\cdots,d_{K_R-1}]$. For a specific demand vector, since the receivers have already cached some packets of their requested files, the transmitters only need to deliver the remaining packets to those receivers. The task in this phase is to design an efficient transmission procedure based on the cache placement in the prefetching phase so that the receivers' demands can be satisfied. In order to guarantee 
that any possible demands can be satisfied, we require that the entire file library is cached among all transmitters,
i.e., $K_TM_T\geq N$.

{For each cached packet $w_{n,p}\in \mathbb{F}_2^{L}$, the transmitter performs random Gaussian coding scheme $\psi:\mathbb{F}_2^{L}\mapsto \mathbb{C}^{\hat{L}}$ with rate $\log P +o(\log P)$ to obtain the coded packet $\hat{w}_{n,p}\triangleq \psi(w_{n,p})$ consisting of $\hat{L}$ complex symbols, so that each coded packet carries  one DoF.}  Assume that the communication will take place in $H$ blocks, each of which consists of $\hat{L}$ time slots. In addition,
we allow only one-shot linear transmission schemes
in each block $m\in[1:H]$ to deliver a set of requested (coded) packets $\mathcal{P}_m$ to a subset of the receivers, denoted by $\mathcal{R}_m$. That is, each transmitter Tx$_i, i\in [K_T]$ will send a linearly coded message 
\begin{equation}\label{linear combo coefficients}
s_i^m=\sum_{(n,p):w_{n,p}\in\mathcal{C}^{\rm T}_i\cap\mathcal{P}_m}\alpha_{i,n,p}^m\hat{w}_{n,p},
\end{equation} 
where $\mathcal{C}^{\rm T}_i$ denotes the cached contents of Tx$_i$ and $\alpha_{i,n,p}^m$ is the linear combination coefficients used by Tx$_i$ at the $m$-th block. Accordingly, the received signal of the intended receivers Rx$_j,j\in\mathcal{R}_m$ in the $m$-th block is
\begin{equation}
y_j^m=\sum_{i=0}^{K_T-1}h_{ji}s_i^m + n_j^m,
\end{equation}  
where $n_j^m\in\mathbb{C}^{\hat{L}}$ is the random
noise at Rx$_j$ in block $m$. Each receiver will utilize its cached contents, consisting of packets stored in the prefetching phase, to subtract some of the interference caused by undesired packets. In particular, each receiver will perform a linear combination $\mathcal{L}_j^m(.)$ if possible in block $m$ 
to recover its requested packets from all received signals as follows  
\begin{equation}
\label{eq: point-2-point channel}
\mathcal{L}_j^m(y_j^m, \hat{\mathcal{C}}_j^{\rm R})=\hat{w}_{d_j,p} + n_j^m,
\end{equation} 
where $\hat{w}_{d_j,p}\in\mathcal{P}_m$ is the desired coded packet of Rx$_j$ and $\hat{\mathcal{C}}_j^{\rm R}$ denotes the Gaussian coded version of the packets cached by Rx$_j$. {The channel created by (\ref{eq: point-2-point channel}) is a point-to-point channel with capacity  $\log P + o(\log P)$. Since each coded packet $\hat{w}_{d_j,p}$ is encoded with rate  $\log P + o(\log P)$, it can be decoded with vanishing error probability as $L$ increases. }

{Since each coded packet carries exactly one DoF, a sum-DoF of $|\mathcal{P}_m|$ can be achieved in block $m$. Therefore, the one-shot linear sum-DoF of $\left| \cup_{m=1}^H \mathcal{P}_m\right|/H $ can be achieved throughout the delivery phase}. As a result, the \emph{one-shot linear sum-DoF} is defined as the maximum achievable one-shot linear sum-DoF for the worst-case demands under a given caching realization \cite{Naderializadeh2017interference}, i.e.,
\begin{equation}
\mathsf{DoF}_{\rm L,sum}^{\left(\left\{\mathcal{C}_i^{\rm T}\right\}_{i=0}^{K_T-1},\left\{\mathcal{C}_j^{\rm R}\right\}_{j=0}^{K_R-1}\right)}=\inf\limits_{\mathbf{d}}\sup\limits_{H,\left\{\mathcal{P}_m\right\}_{m=1}^H}\frac{\left|\bigcup_{m=1}^H\mathcal{P}_m\right|}{H}.
\end{equation}
The \emph{one-shot linear sum-DoF of the network} is correspondingly defined as the maximum achievable one-shot linear sum-DoF over all possible caching realizations, i.e.,
\begin{equation}
\begin{aligned}
&\mathsf{DoF}_{\rm L,sum}^{\ast}(N,M_T,M_R,K_T,K_R)=
\sup\limits_{\left\{\mathcal{C}_i^{\rm T}\right\}_{i=0}^{K_T-1},\left\{\mathcal{C}_j^{\rm R}\right\}_{j=0}^{K_R-1}}\mathsf{DoF}_{\rm L,sum}^{ \left(\left\{\mathcal{C}_i^{\rm T}\right\}_{i=0}^{K_T-1},\left\{\mathcal{C}_j^{\rm R}\right\}_{j=0}^{K_R-1}\right)},
\end{aligned}
\end{equation}
in which the cached contents of all transmitter and receivers satisfy the memory constraints, i.e., $|\mathcal{C}_i^{\rm T}|\leq M_TF,\forall i\in[K_T]$ and $|\mathcal{C}_j^{\rm R}|\leq M_RF,\forall j\in[K_R]$.

\subsection{Combinatorial Cache Placement Design}
\label{subsec:Combinatorial Cache Placement Design}
In this paper, the combinatorial cache placement design based on {\em hypercube}, proposed in \cite{woolsey2020d2d,woolsey2020new} to reduce the subpacketization level in wireless D2D networks is adopted in the prefetching phase. The hypercube cache placement 
has a nice geometric interpretation: each packet of the file can be represented by a lattice point in a high-dimensional hypercube and the cached content of each D2D node is represented by a hyperplane in that hypercube (see Fig.~\ref{fig: HC}). Based on the hypercube cache placement and the corresponding communication scheme, order-optimal rate can be achieved with exponentially less number of packets compared to the Ji-Caire-Molisch (JCM) scheme \cite{ji2016fundamental}.
It turns out that by a non-trivial extension, the hypercube scheme can also significantly reduce the required subpacketizations 
in cache-aided interference networks. 
The details of hypercube cache placement \cite{woolsey2020d2d,woolsey2020new} is described as follows. 

\subsubsection{Hypercube cache placement design for wireless D2D caching networks}
Consider a wireless D2D network consisting of a library of $N$ files, each with $F$ packets, and $K$ users, each of which is equipped with a local cache memory of size $M$ files, or equivalently, $MF$ packets. The \emph{caching parameter}, defined as $t\triangleq {KM}/{N}\in[1:K]$, represents the average number of times that each file is cached among all users. 
In the hypercube cache placement, each file $\mathcal{W}_n$ is split into $\left({N}/{M}\right)^t$ subfiles\footnote{{{In the prefetching phase, each file is split into multiple smaller files and each of these smaller subfiles is then spread across the user caches. We use ``subfile'' to refer to these smaller subfiles. In the delivery phase, in order to perform interference cancellation, each subfile needs to be further split into multiple even smaller ones. We use ``packet'' to refer to such smaller files resulting from splitting the subfiles. So a packet is the smallest unit that a file is split into. To transmit these requested packets to the target receivers, random Gaussian encoding must be applied to these packets and the output is called \emph{coded packets}. However, in the description of the general schemes, we just refer the coded packet to as packet for simplicity when there is no confusion.}}} (assuming that ${N}/{M}$ and $t$ are both positive integers), i.e., $\mathcal{W}_n=\left\{\mathcal{W}_{n, (\ell_0,\ell_1,\cdots,\ell_{t-1})}: \,\ell_j\in[{N}/{M}],\,j\in[t]\right\}$. It can be seen that each subfile of a file $\cw_n$ is uniquely marked by a $t$-tuple $(\ell_0,\ell_1,\cdots,\ell_{t-1})$ where $\ell_j,j\in[t]$ represents the index of the lattice point along the $j$-th dimension. In the prefetching phase, each user $u\in[K]$ caches a set of subfiles $\left\{\mathcal{W}_{n,(\ell_0,\ell_1,\cdots,\ell_{t-1})}: \,\forall n\in[N]\right\}$, where $\ell_j=u\mod ({N}/{M})$, for $j=\lfloor u/\left({N}/{M}\right)\rfloor$, and $\ell_i\in[{N}/{M}]$ for any $i\neq j$. As a result, each user will cache $({N}/{M})^{t-1}$ subfiles from each file $\mathcal{W}_n$. It can be verified that the total number of subfiles cached by any user is equal to $N(\frac{N}{M})^{t-1} =N\frac{(N/M)^t}{N/M}= N \frac{F}{N/M} = MF$, satisfying the memory constraint. The hypercube cache placement has a 
nice geometric interpretation. 
Under the hypercube file splitting method, each subfile will represent a lattice point with coordinate $(\ell_0,\ell_1,\cdots,\ell_{t-1})$ in a $t$-dimensional hypercube, and ${N}/{M}\in\mbb{Z}^+$ is the number of lattice points along each dimension. 
We will further illustrate the details of the hypercube cache placement via the following example. 

\begin{figure}  
\centering
\includegraphics[width=0.60\textwidth]{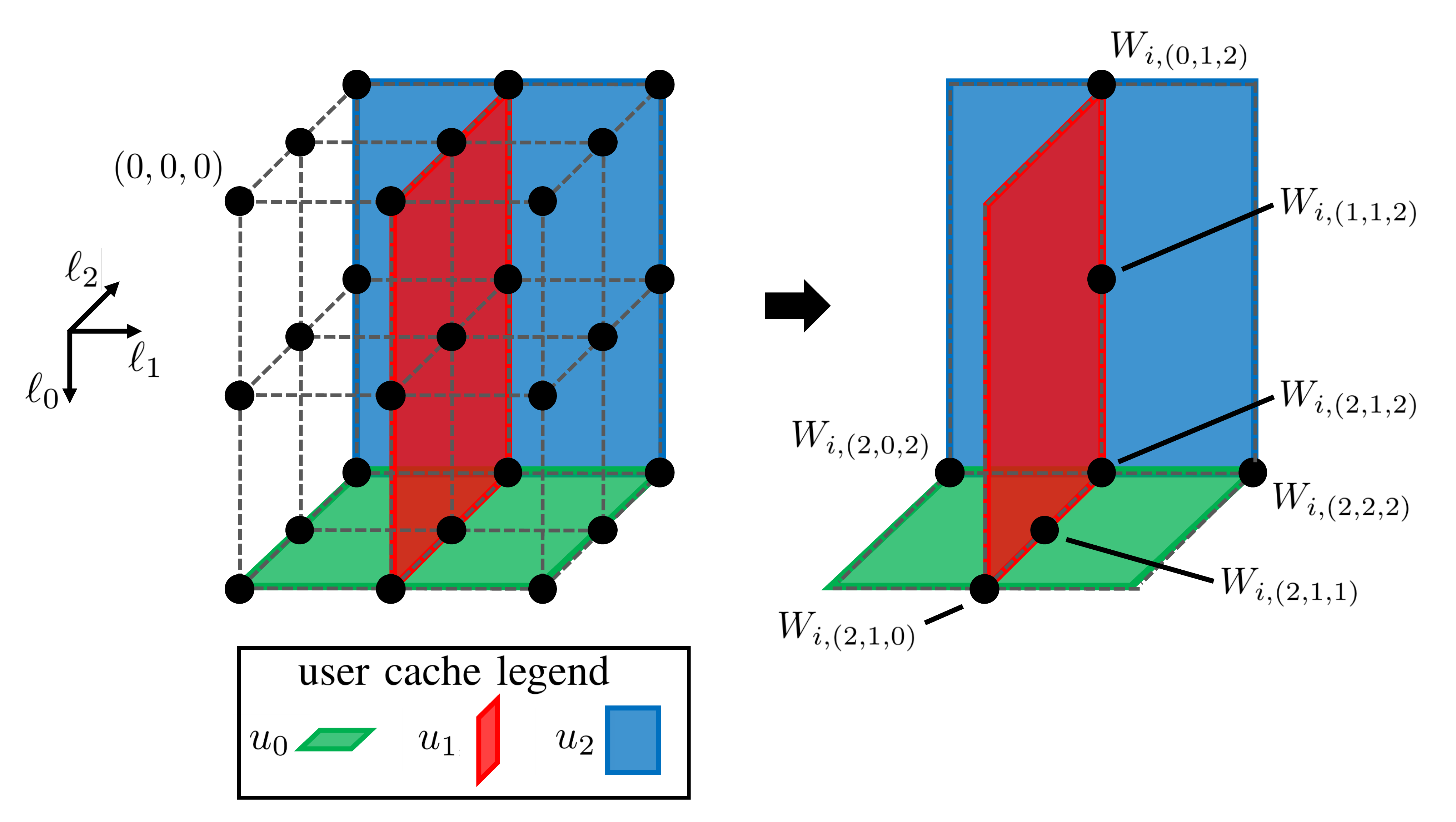}
\caption{A 3-dimensional example of the hypercube cache placement. Each subfile is represented by a unique lattice point in the 3-dimensional hypercube (cube). Each of the 9 users caches a set of packets represented by plane of lattice points. As a result, each user caches $9\times 9=81$ subfiles in total.}
\label{fig: HC}
\label{figure:2}

\end{figure}

\begin{example} \tbf{(Hypercube Cache Placement)}
Consider a set of $K=9$ users labeled as $0,1,\cdots, 8$ and a set of $N=9$ files $\{\Wc_n,  n \in[9]\}$. Each user has a cache memory of size $M=3$ files. We first partition the users into $t\triangleq {KM}/{N}=3$ groups denoted by $\mathcal{U}_0=\{0,1,2\}$, $\mathcal{U}_1=\{3,4,5\}$ and $\mathcal{U}_2=\{6,7,8\}$. 
Each file $\mathcal{W}_n$ is split into $\left({N}/{M}\right)^t=27$ subfiles, i.e., $\mathcal{W}_n=\{\mathcal{W}_{n,(\ell_0,\ell_1,\ell_2)}:\ell_0,\ell_1,\ell_2\in[3]\}$, each of which can be represented by a unique lattice point in the 3-dimensional cube (see Fig. \ref{figure:2}). As a result, each lattice point will represent a set of $N=9$ subfiles, each from a distinct file. For the cache placement, each user caches all subfiles represented by a plane of lattice points of the cube. For example, user $u_0=2,u_1=4$ and $u_2=8$ will cache subfiles represented by the green, red and blue planes respectively in Fig. \ref{figure:2}. We can see that the set of subfiles $\{\mathcal{W}_{n,(2,1,2)}:\forall n \in [9]\}$ represented by the lattice point $(2,1,2)$, which is the intersection of the three orthogonal planes of different colors, is cached exclusively by users $u_0,u_1$ and $u_2$. Similarly, each subfile is cached by three distinct users.  
\hfill $\triangle$
\end{example}

\subsubsection{Hypercube cache placement design for cache-aided interference networks}
\label{subsubsec:Hypercube cache placement design for cache-aided interference networks}
Different from the D2D setting in \cite{woolsey2020d2d}, in cache-aided interference networks, we have  a set of explicit transmitters and receivers instead of D2D users. However,
the hypercube approach can still be applied to design the cache placement in the case illustrated as follows. 

\tit{File Splitting:} let $D_T\triangleq {N}/{M_T} \in \mathbb{Z}^+$ and $D_R\triangleq {N}/{M_R} \in \mathbb{Z}^+$ denote the number of transmitters and receivers on each edge of the hypercube associated with the transmitters' cache and receivers' cache respectively.\footnote{{Since we apply the hypercube cache placement at both the transmitters' and receivers' sides, there are two hypercubes associated with the cache-aided interference network, including the \tit{transmitter hypercube} which is a $t_T$-dimnesional hypercube with each edge containing $N/M_T$ lattice points (transmitters), and the \tit{receiver hypercube} which is a $t_R$-dimnesional hypercube with each edge containing $N/M_R$ lattice points (receivers).}}  For the set of $K_T=D_Tt_T$ transmitters $\{\textrm{Tx}_k: k\in[K_T]\}$, we denote the $t_T\triangleq {K_TM_T}/{N}$ dimensions of the transmitters as $\mathcal{U}^{\rm T}_i=\{k:\lfloor {k}/{D_T} \rfloor=i\},\forall i \in [t_T]$\footnote{{The superscript ``T'' means ``Transmitter''. Readers should not confuse this with the transpose operator.}}. Similarly, for the set of $K_R=D_Rt_R$ receivers $\{\textrm{Rx}_k: k\in[K_R]\}$, we denote the $t_R \triangleq {K_RM_R}/{N}$ dimensions of the receivers as $\mathcal{U}^{\rm R}_j=\{k:\lfloor {k}/{D_R} \rfloor=j\},\forall  j \in [t_R]$. It can be seen that $|\mathcal{U}^{\rm T}_i|=D_T,\,\forall i\in[t_T]$ and $|\mathcal{U}^{\rm R}_j|=D_R,\,\forall j\in[t_R]$, i.e., for both the transmitter and the receiver hypercubes, all
distinct dimensions (edges) contain the same number of lattice points. With this file splitting, the prefetching phase is then described as follows.

\textit{Prefetching Phase}: The hypercube cache placement is employed at both the transmitters' and receivers' sides. That is, each file $\mathcal{W}_n$ is split into ${D_T}^{t_T}{D_R}^{t_R}=(\frac{N}{M_T})^{t_T}(\frac{N}{M_R})^{t_R}$ disjoint equal-size subfiles, denoted by
\begin{equation}
\mathcal{W}_n=\left\{\mathcal{W}_{n,\mathcal{T},\mathcal{R}}\right\}_{\substack{\mathcal{T}\in \mathcal{U}^{\rm T}_0\bigotimes\mathcal{U}^{\rm T}_1\bigotimes\cdots\bigotimes\mathcal{U}^{\rm T}_{t_T-1}\\
\mathcal{R}\in \mathcal{U}^{\rm R}_0\bigotimes\mathcal{U}^{\rm R}_1\bigotimes\cdots\bigotimes\mathcal{U}^{\rm R}_{t_R-1}}},
\end{equation}
in which the definition of the operator $\bigotimes$ is as follows. For $m\in\mbb{Z}^+$ sets $\mc{A}_0,\mc{A}_1,\cdots, \mc{A}_{m-1}$, we define $
\mc{A}_0\bigotimes\mc{A}_1\bigotimes\cdots \bigotimes\mc{A}_{m-1}$ as the set of all un-ordered elements in $\mc{A}_0\times\mc{A}_1\times\cdots \times\mc{A}_{m-1}$, where $\times$ denotes the Cartesian product. We use $\{\cdot\}$ to convert the $m$-tuple $A$ to a set. For example, for a tuple $(1,2,3)$, we have $\{(1,2,3)\}=\{1,2,3\}$. Hence, $\mc{A}_0\bigotimes\mc{A}_1\bigotimes\cdots \bigotimes\mc{A}_{m-1}\triangleq \left\{ \{A\}: A\in \mc{A}_0\times\mc{A}_1\times\cdots \times\mc{A}_{m-1}   \right\}$.
The subfile $\mc{W}_{n,\mc{T},\mc{R}}$ is exclusively cached by a set of transmitters in $\mc{T}$ and a set of receivers in $\mc{R}$. Under this file splitting strategy, each transmitter Tx$_i$ caches a set of subfiles $\{\mathcal{W}_{n,\mathcal{T},\mathcal{R}}:\forall \mathcal{T}:i\in\mathcal{T},\forall \mathcal{R},\forall n\in[N]\}$ and each receiver Rx$_j$ caches a set of subfiles $\{\mathcal{W}_{n,\mathcal{T},\mathcal{R}}: \forall \mathcal{T},\forall \mathcal{R}:j\in\mathcal{R},\forall n\in[N]\}$. As a result, the number of subfiles cached by Tx$_i,i\in[\kt]$ is equal to $N\dt^{\ttt-1}\dr^\ttr$ and hence the number of packets cached by Tx$_i,i\in[\kt]$ is equal to 
\be
\label{eq: Tx memory}
N{D_T}^{t_T-1}{D_R}^{t_R}\frac{F}{{D_T}^{t_T}{D_R}^{t_R}}=M_TF, 
\ee
where $\frac{F}{{D_T}^{t_T}{D_R}^{t_R}}$ is the number of packets of each subfile (note that in the following delivery phase, each subfile needs to be further split into multiple packets). 
Similarly, the number of subfiles cached by Rx$_j,j\in[\kr]$ is equal to $N\dt^{\ttt}\dr^{\ttr-1}$ and hence the number of packets cached by Rx$_j,\forall j\in[\kr]$ is equal to 
\be N{D_T}^{t_T}{D_R}^{t_R-1}\frac{F}{{D_T}^{t_T}{D_R}^{t_R}}=M_RF,
\ee which also satisfies the memory constraint. The application of the hypercube cache placement method to cache-aided interference networks is illustrated via the following example. 

\begin{example} (\textbf{Hypercube Cache Placement for Interference Networks})
\label{ex: prefetching interference}
Consider a wireless network with $K_T=4$ transmitters and $K_R=4$ receivers. Each transmitter and receiver is equipped with a cache memory of size $M_T=2$ and $M_R=2$ files, respectively. The file library contains $N=4$ files denoted by  $A,B,C$ and $D$. Hence, we have the parameters $\dt={N}/{\mt}=2, \dr={N}/{\mr}=2$, $\ttt={\kt}/{\dt}=2$ and $\ttr={\kr}/{\dr}=2$. In this case, both the transmitter and receiver hypercubes are two-dimensional hypercubes (i.e., squares) with each edge containing two transmitters/receivers.
 
In the prefetching phase, each file $\mathcal{W}_n$ is split into $\dt^\ttt\dr^\ttr=16$ 
subfiles $\{\mathcal{W}_{n,\mathcal{T},\mathcal{R}}\}$ of equal sizes for any $\mathcal{T}\in\left\{\{0,2\},\{0,3\},\{1,2\},\{1,3\}\right\}$ and $\mathcal{R}\in\left\{\{0,2\},\{0,3\},\{1,2\},\{1,3\}\right\}$. Each subfile is then cached by the two transmitters in $\mathcal{T}$  and the two receivers in $\mathcal{R}$, respectively. For example, file $A$ is split into 16 subfiles:\footnote{{With a slight abuse of notation, we write $A_{\{0, 2\},\{0, 2\}}$ as $A_{02,02}$ for simplicity and the same for other symbols.}}
\begin{eqnarray}
&A_{02,02},\;A_{02,03},\;A_{02,12},\;A_{02,13},\nonumber\\
&A_{03,02},\;A_{03,03},\;A_{03,12},\;A_{03,13},\nonumber\\
&A_{12,02},\;A_{12,03},\;A_{12,12},\;A_{12,13},\nonumber\\
&A_{13,02},\;A_{13,03},\;A_{13,12},\;A_{13,13}, \;\nonumber
\end{eqnarray}
where for example, $A_{02,02}$ is cached by transmitters Tx$_0$ and Tx$_2$ as well as receivers Rx$_0$ and Rx$_2$. The same file splitting is done for files $B,C$ and $D$. 
It can be seen that each transmitter caches 8 subfiles of each file. Since each subfile contains ${F}/{16}$ packets, the total number of packets cached by each transmitter is $4\times 8\times {F}/{16}=2F$, which satisfies the memory constraint of the transmitters. Similarly, the memory constraint of the receivers is also satisfied. \hfill $\triangle$
\end{example}

\section{Main Result}
\label{section: main result}
The main results on the one-shot linear sum-DoF using the hypercube cache placement approach are presented in this section.
{Note that when $\frac{K_TM_T+K_RM_R}{N}> K_R$, the sum-DoF $K_R$ is always achievable by only utilizing a fraction of the Tx/Rx cache memories such that for the updated system with Tx/Rx cache memories $M_T'\le M_T$ and $M_R'\le M_R$, we have $\frac{K_TM_T'+K_RM_R'}{N}=K_R$. Therefore, by applying the proposed scheme on the updated system, the sum-DoF of  $K_R$ can be achieved. As a result, we focus on the case where $\frac{K_TM_T+K_RM_R}{N}\le K_R$.}
\begin{theorem}
\label{theorem:1}
For a $K_T\times K_R$ wireless interference network with a library of $N$ files, each consisting of $F$ packets, and with transmitter and receiver cache sizes of $M_TF$ and $M_RF$ packets,  respectively, given the hypercube cache placement approach employed in the prefetching phase, and for any $\delta\triangleq {t_T}/{t_R}\in\mathbb{Z}^+$, $D_R=N/M_R\geq \delta +1$, where {$t_T\in[1:K_T],t_R\in[K_R],\dr\in\mbb{Z}^+$}, the one-shot linear sum-DoF of $\frac{K_TM_T+K_RM_R}{N}$ is achievable when $K_R \geq \frac{K_TM_T+K_RM_R}{N}$ with 
\be
F = \left(\frac{N}{M_T}\right)^{t_T}\left(\frac{N}{M_R}\right)^{t_R}\binom{D_R-2}{\delta-1}{\binom{D_R-1}{\delta}}^{t_R-1}\frac{(\delta!)^{t_R}}{\delta}(t_R-1)!
\ee 
\end{theorem}
\begin{IEEEproof}
The achievability of Theorem \ref{theorem:1} is proved by the general achievable scheme  described in Section \ref{subsec:General Achievable Scheme}, which focuses on the case $\kr\ge \tron$. The converse results follows directly from \cite{Naderializadeh2017interference} which will not be presented in this paper.
\end{IEEEproof}
    
The implications of Theorem \ref{theorem:1} are two-folds, which includes the optimality of the achievable one-shot linear DoF and the reduced subpacketization level. 
Note that if either $\ttt$ or $\ttr$ is not an integer, or both of them are not integers, we can still achieve the sum-DoF of $\ttt+\ttr$ for any values of $\ttt$ and $\ttr$ using the 
the \emph{memory-sharing} in \cite{maddah2014fundamental} which will be briefly introduced later. The following observations are ready.

\subsubsection{Sum-DoF Optimality} As 
shown in \cite{Naderializadeh2017interference}, when $K_R \geq \frac{K_TM_T+K_RM_R}{N}$, the optimal one-shot linear sum-DoF of the interference network studied in this  paper, $\msf{DoF}^*_{\rm L,sum}$, over any possible cache placement realizations, is bounded by 
$
\frac{\scriptstyle K_TM_T+K_RM_R}{\scriptstyle N} \leq\mathsf{DoF}_{\rm L,sum}^{\ast}\leq \frac{\scriptstyle 2(K_TM_T+K_RM_R)}{\scriptstyle N},
$
which indicates that when $\frac{K_TM_T+K_RM_R}{N}\le \kr$, the achievable one-shot linear sum-DoF under the hypercube cache placement is equal to the achievable one-shot linear DoF in \cite{Naderializadeh2017interference} and is within a factor of $2$ to the optimal one-shot linear sum-DoF of the network. This result indeed shows that the DoF of $\tron$ can be achieved by different cache placement methods, which provides the potential to reduce the total number of packets required. 
{In addition,  to see how much DoF gain can be obtained going beyond one-shot linear transmissions, we refer the readers to Section VI of \cite{Naderializadeh2017interference} where the scaling law of the optimal sum-DoF is analyzed. In particular, for the cases of large number of transmitters and receivers ($K_T=K_R=K \rightarrow \infty$ and other parameters are fixed) and constant number of transmitters $(K_T=C,K_R=K \rightarrow \infty)$, the one-shot linear scheme achieves the same DoF scaling as the interference alignment alike schemes. This implies that interference alignment alike multi-shot schemes can only provide constant DoF gain over the one-shot linear schemes which has much lower complexity.}


\subsubsection{Subpacketization Level Reduction} Under the hypercube cache placement strategy, the number of packets per file, i.e., $F$, required for implementing the interference cancellation in the delivery phase is significantly reduced compared to the NMA scheme. In particular, the NMA scheme requires to split each file into $\binom{K_T}{t_T}\binom{K_R}{t_R}$ subfiles in the prefetching phase and further split each subfile into $\frac{t_R![K_R-(t_R+1)]!}{[K_R-(t_T+t_R)]!}$  packets in the delivery phase. However, if we employ the hypercube cache placement strategy, each file is going to be split into $(\frac{N}{M_T})^{t_T}(\frac{N}{M_R})^{t_R}$ subfiles in the prefetching phase, and is further split into $\binom{D_R-2}{\delta-1}{\binom{D_R-1}{\delta}}^{t_R-1}\frac{(\delta!)^{t_R}}{\delta}(t_R-1)!$ packets\footnote{{Here we have implicitly assumed that $D_R-2\geq \delta-1$, i.e., $D_R\geq \delta+1$. This assumption can be justified as follows. In real-world wireless networks, the number of receivers (users) $K_R$ can be larger than the number of transmitters (base stations, BS) $K_T$, since each BS can be associated with multiple users. However, each BS can have large cache memory than the users, i.e., $M_T \gg M_R$. Due to the larger per-BS cache memory $M_T$ but relatively small $K_T$ at the transmitter's side and the smaller per-user cache memory $M_R$ but larger $K_R$ at the users' sides, it is reasonable to assume that the caching parameters $t_T$ and $t_R$ are close to each other. Also, since each user's cache memory is very small compared to the file library, i.e., $M_R/N \ll 1$, then $D_R=N/M_R \gg 1$. For example, consider a network with $N=500$ files each having size 5 GB (e.g., Netflix movies), $K_T=5$ BSs each capable of caching $M_T=200$ files (i.e., 1000 GB memory per BS), and $K_R=50$ receivers each capable of caching $M_R=10$ files (50 GB memory per receiver). In this case, we have $t_T=K_TM_T/N=5\times 200/500=2, t_R=K_RM_R/N=50\times 10/500=1$ and therefore $\delta=1$. Moreover, we have $D_R=N/M_R=500/10=50$ which is much larger than $\delta+1=2$. As a result, it can be seen that the assumption $D_R\ge \delta +1$ is valid
in practice.}}
in the delivery phase. In Section \ref{complexity analysis}, we will show that for any system parameters, the hypercube scheme requires less number of packets than the NMA scheme and the gain of subpacketization can be unbounded with the increase of the cache sizes of transmitters and receivers. Together with the sum-DoF optimality, the hypercube
based scheme can achieve the same one-shot linear DoF as in \cite{Naderializadeh2017interference} while requiring a significantly 
smaller $F$.

\subsubsection{Non-integer Caching Parameters $t_T,t_R$} When the caching parameters $\ttt=\frac{\kt\mt}{N}$ and/or $\ttr=\frac{\kr\mr}{N}$ are not integers, we can still achieve the one-shot linear sum-DoF of $\ttt+\ttr$ using the \emph{memory-sharing} method of \cite{maddah2014fundamental}. More specifically, we can split the Tx/Rx memories and files proportionally so that for each of the new partitions, our proposed scheme can be applied for the updated parameters $\ttt'$ and $\ttr'$ which are integers. That is, for each new partition of memories and files, it can be treated as a new interference network with updated Tx/Rx cache memories $M_T',M_R'$, file size $L'$ and the corresponding caching parameters $t_T'=\frac{\kt M_T'}{N}\in\mathbbm{Z}^+,t_R'=\frac{\kr M_R'}{N}\in\mathbbm{Z}^+$, where the proposed scheme can be directly applied. 

{\subsubsection{Non-integer Values of $\delta$} Although in Theorem 1 we have assumed that $\delta=\frac{t_T}{t_R}\in\mathbb{Z}^+$, the sum-DoF of $t_T+t_R$ can also be achieved even when $\delta$ is not an integer. This can be done following a similar method to memory sharing. Note that $\delta \ge 1/t_R$ due to $t_T\ge 1$ since the file library has to be stored at least once by all the transmitters otherwise the receivers' demands can not be satisfied. Now we consider the case when both $t_T$ and $t_R$ are positive integers but $\delta=\frac{t_T}{t_R}$ is not an integer. The scheme to achieve the sum-DoF $t_T+t_R$ is described as follows.

We can split the Tx/Rx cache memories and the files proportionally such that the updated caching parameters $t_T'$ and $t_T''$ of each partition correspond to $\delta'$ and $\delta''$ both of which are integers. More specifically, each Tx memory is split into two parts $M_T'=\alpha M_T$ and $M_T''=(1-\alpha) M_T$ for some $0<\alpha<1$, each Rx memory is split into two parts $M_R'=\beta M_R$ and $M_R''=(1-\beta) M_R$ for some $0<\beta<1$, and each file $\mathcal{W}_n$ is split into two parts $\mathcal{W}_n=(\mathcal{W}_n' ,\mathcal{W}_n'')$ where $|\mathcal{W}_n'|=\gamma|\mathcal{W}_n|$ and $|\mathcal{W}_n''|=(1-\gamma)|\mathcal{W}_n|$ for some $0<\gamma<1$. We then apply the proposed scheme on the two Tx/Rx memory and file partitions $(M_T',M_R',\{\mathcal{W}_n'\}_{n\in[N]}, t_T',t_R')$  and $(M_T'',M_R'',\{\mathcal{W}_n''\}_{n\in[N]},t_T'',t_R'')$ where $\delta'=t_T'/t_R'$ and $\delta''=t_T''/t_R''$ are both integers. WLOG, we let $\beta=\gamma$. Therefore, we have $t_T'=\frac{K_TM_T'}{N}=\frac{\alpha}{\gamma}t_T, t_R' =\frac{K_RM_R'}{N}=t_R$ and $t_T''=\frac{K_TM_T''}{N}=\frac{1-\alpha}{1-\gamma}t_T, t_R'' =\frac{K_RM_R''}{N}=t_R$ and $t_T=pt_T' + (1-p)t_T'', \delta =p\delta'+(1-p)\delta''$ where $p=\frac{t_T''-t_T}{t_T''-t_T'}$. Next we consider two different cases: \emph{1)} $\delta\in[1/t_R,1)$, and \emph{2)} $\delta \in(q,q+1)$ for some $q\in\mathbb{Z}^+$.

$\bullet $ \emph{Case 1}: $\delta\in[1/t_R,1)$. Let $t_T'=1,t_T''=t_R$. For the first memory and file partition, the coded caching scheme \cite{maddah2014fundamental} can be applied to achieve the sum-DoF of $t_R'+1=t_R+1$; For the second memory and file partition,  we can apply the proposed scheme with $\delta''=1$ to achieve the sum-DoF $t_T''+t_R''=2t_R$. As a result, the overall sum-DoF of $p(t_R'+1)+(1-p)(t_T''+t_R'')=t_T+t_R$   can be achieved.

$\bullet $ \emph{Case 2}: $\delta \in(q,q+1)$ for some $q\in\mathbb{Z}^+$. Let $t_T' = qt_R=\lfloor \delta \rfloor t_R$ and $t_T''=(q+1)t_R =\lceil \delta \rceil t_R $. For  the first and second memory and file partitions with $\delta'=\frac{t_T'}{t_R'}=\lfloor \delta \rfloor $ and  $\delta''=\frac{t_T''}{t_R''}=\lceil \delta \rceil $, the proposed scheme can be directly applied to achieve the sum-DoF of $t_T'+t_R'=(\lfloor \delta \rfloor +1)t_R$ and $t_T''+t_R''=(\lceil \delta \rceil +1)t_R$ respectively. Therefore, the overall  sum-DoF $p(t_T'+t_R')+(1-p)(t_T''+t_R'') = t_T+t_R$ can be achieved. In both cases, let $F'$ and $F''$ be the required number of packets per file over the two memory and file partitions which can be calculated by Eq. (21). Then the number of packets per file is determined as $F=F'+F''$. }

\section{Achievable Delivery Scheme}
\label{section: achievable scheme}

\subsection{An Example}
\begin{figure*}
\centering
\includegraphics[width=0.7\textwidth]{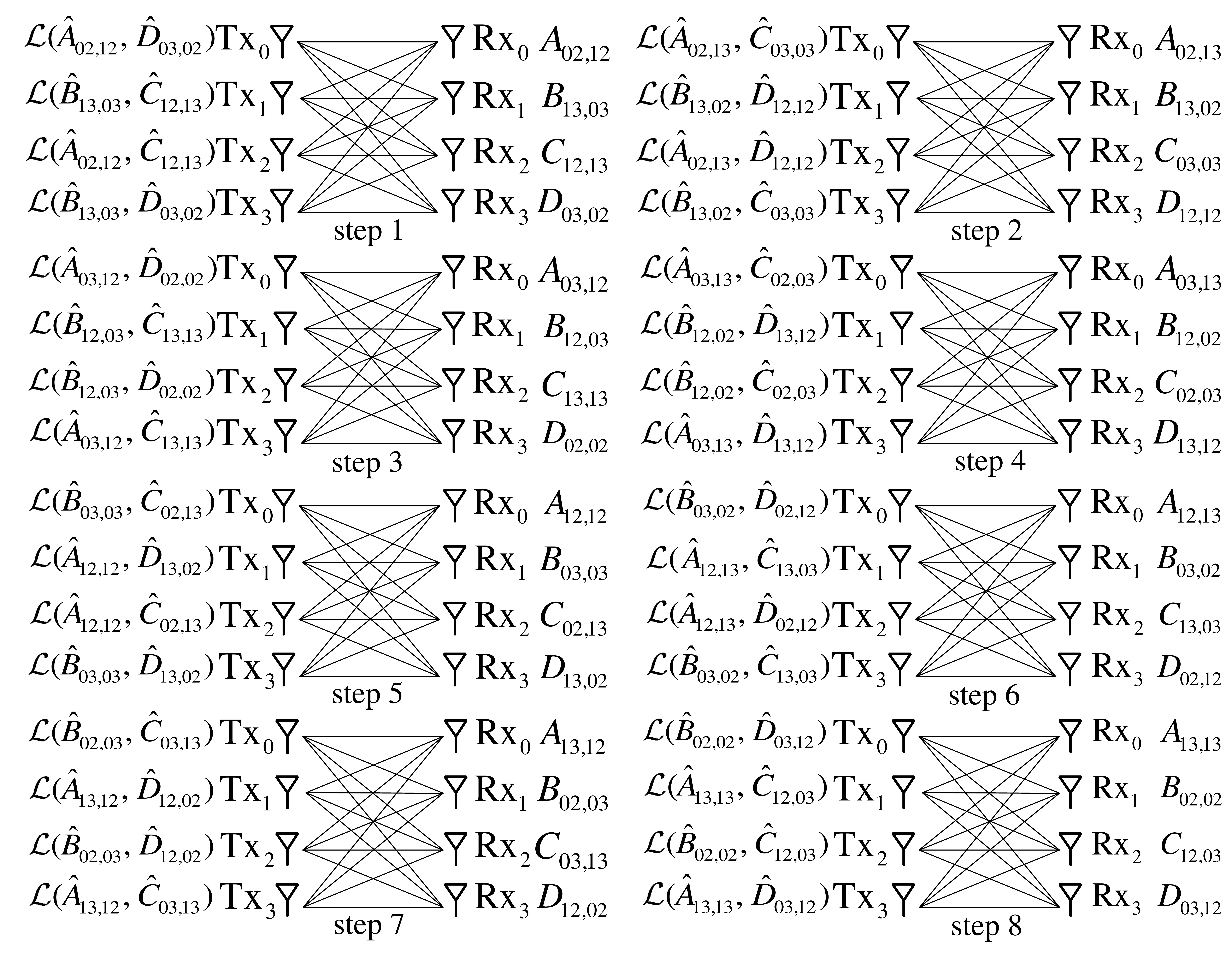}
\caption{Delivery phase for Example \ref{ex: delivery interference} in which four receivers Rx$_j,j \in [4]$ request four different files $A,B,C$ and $D$ respectively. $\mathcal{L}({x,y})$ denotes some linear combination of $x$ and $y$, i.e., $\mathcal{L}(x,y)=\alpha x+\beta y$, where $\alpha$ and $\beta$ are some constants. There are in total 8 communication steps and in each of which 4 different packets are delivered to the receivers interference-free.} 
\label{figure:3}
\end{figure*}

We first present the achievable delivery scheme under the hypercube cache placement via the following example. 
\begin{example}
\label{ex: delivery interference} 
\tbf{(Achievable Delivery Scheme)} We consider the same network setting as in Example \ref{ex: prefetching interference}. Let 
receiver Rx$_j$ request 
the file $\mathcal{W}_{d_j}$. 
Without loss of generality, we assume that $\mathcal{W}_{d_0}=A,\mathcal{W}_{d_1}=B,\mathcal{W}_{d_2}=C$ and $\mathcal{W}_{d_3}=D$. In the prefetching phase, each receiver has already cached 8 subfiles of its requested file. Therefore, the transmitters only need to deliver the $16-8=8$ remaining subfiles to each receiver. 
In particular, the following $32$ subfiles need to be delivered to the receivers:
\[
 \begin{array}{ccc}
A_{02,12},\;A_{03,12},\;A_{12,12},\;A_{13,12},\\
A_{02,13},\;A_{03,13},\;A_{12,13},\;A_{13,13}\:\\
\end{array}
\Big\}\:\textrm{to Rx}_0,\quad   \begin{array}{ccc}
B_{02,02},\;B_{03,02},\;B_{12,02},\;B_{13,02},\\
B_{02,03},\;B_{03,03},\;B_{12,03},\;B_{13,03}\:\\
\end{array}
\Big\}\:\textrm{to Rx}_1,   
\]
\[
 \begin{array}{ccc}
C_{02,03},\;C_{03,03},\;C_{12,03},\;C_{13,03},\\
C_{02,13},\;C_{03,13},\;C_{12,13},\;C_{13,13}\:\\
\end{array}
\Big\}\:\textrm{to Rx}_2,\quad \begin{array}{ccc}
D_{02,02},\;D_{03,02},\;D_{12,02},\;D_{13,02},\\
D_{02,12},\;D_{03,12},\;D_{12,12},\;D_{13,12}\:\\
\end{array}
\Big\}\:\textrm{to Rx}_3.   
\]

Note that in the hypercube-based delivery scheme, each subfile needs to be further split into 
$\fr$ 
packets. 
In this example, since $\delta=\frac{\ttt}{\ttr}=1,\dt = \ttt = \dr=\ttr=2$, $\delta=1$, we have $
\fr=\binom{0}{0}\binom{1}{1}(2-1)!=1 $, implying that no further file splitting is needed
and thus $32$ packets will be delivered. 

We now show how the above $32$ packets can be grouped in $8$ subsets, each of which contains $4$ packets, such that the packets within the same subset can be delivered simultaneously to the receivers without interference. 
Fig. \ref{figure:3} shows how the 32 packets to be delivered are grouped and transmitted. In each communication step, $\ttt+\ttr = 4$ packets are delivered to the receivers simultaneously, and the interference among different users can be effectively eliminated by choosing proper linear combination coefficients at the  $\ttt+\ttr = 4$ transmitters. For example, in step 1 of Fig. \ref{figure:3}, four packets $A_{02,12},B_{13,03},C_{12,13}$ and $D_{03,02}$ are delivered to receivers Rx$_0$, Rx$_1$, Rx$_2$ and Rx$_3$ respectively. We write the transmitted signals $S_i,i\in[4]$ of each transmitter Tx$_i$ as a linear combination of a subset of these four packets as follows:
\begin{eqnarray}
S_0&=&h_{32}\hat{A}_{02,12}-h_{13}\hat{D}_{03,02},\quad 
S_1=h_{23}\hat{B}_{13,03}-h_{02}\hat{C}_{12,13},\nonumber\\
S_2&=&h_{01}\hat{C}_{12,13}-h_{30}\hat{A}_{02,12},\quad 
S_3=h_{10}\hat{D}_{03,02}-h_{21}\hat{B}_{13,03},\nonumber
\end{eqnarray}
where for each packet $\mathcal{W}_{n,\mathcal{T},\mathcal{R}}$, $\hat{W}_{n,\mathcal{T},\mathcal{R}}$ denotes its physical layer coded version. 
As a result, due to the careful choice of the linear coefficients, 
some interference terms are canceled 
over the air by zero forcing (e.g., $\hat{C}_{12,13}$ is canceled at Rx$_0$). The corresponding received signals by Rx$_0$, Rx$_1$, Rx$_2$ and Rx$_3$ after zero forcing are given by
\begin{eqnarray}
Y_0&=&(h_{32}h_{00}-h_{30}h_{12})\hat{A}_{02,12}+(h_{23}h_{01}-h_{21}h_{03})\hat{B}_{13,03}+(h_{10}h_{03}-h_{13}h_{00})\hat{D}_{03,02}+N_0,\nonumber\\
Y_1&=&(h_{23}h_{11}-h_{21}h_{13})\hat{B}_{13,03}+(h_{32}h_{10}-h_{30}h_{12})\hat{A}_{02,12}+(h_{02}h_{11}-h_{01}h_{12})\hat{C}_{12,13}+N_1,\nonumber\\
Y_2&=&(h_{01}h_{22}-h_{02}h_{21})\hat{C}_{12,13}+(h_{32}h_{20}-h_{30}h_{22})\hat{A}_{02,12}+(h_{10}h_{23}-h_{13}h_{20})\hat{D}_{03,02}+N_2,\nonumber\\
Y_3&=&(h_{10}h_{33}-h_{13}h_{30})\hat{D}_{03,02}+(h_{23}h_{31}-h_{21}h_{33})\hat{B}_{13,03}+(h_{01}h_{32}-h_{02}h_{31})\hat{C}_{12,13}+N_3,\nonumber
\end{eqnarray}
where $N_i,i \in [4]$ represents the Gaussian noise.

We can see that receiver Rx$_0$ can cancel the interference caused by $B_{13,03}$ and $D_{03,02}$ since these two packets have already been cached by Rx$_0$ and the desired packet $A_{02,12}$ can be successfully decoded by subtracting the undesired but prefetched packets. Similarly, Rx$_1$, Rx$_2$ and Rx$_3$ can also cancel the interference caused by undesired packets by utilizing their cached contents. Therefore, all the interference including inter-user interference and {interference that can be nulled out by cached packets} can be eliminated so that all receivers can decode their desired packets. It can be verified that there exist such linear combinations and all receivers can decode their desired packets in all remaining $7$ communication steps. Hence, the $32$ packets, each consisting of ${|\Wc_n|}/{16}$ bits, can be delivered to the receivers in 8 communication steps, each containing ${F}/{16} = 1$ resource block. As a result, a sum-DoF of 
$\frac{K_TM_T+K_RM_R}{N}=4$ can be achieved. 
Hence, the proposed file subpacketization, cache placement, 
precoding and scheduling strategy in the delivery phase allow transmitters to collaboratively
zero-force some of the outgoing interference and allow receivers to cancel the leftover interference 
using cached contents for any receivers' demands.
\hfill $\triangle$
\end{example}

\subsection{Hypercube Permutation}
\label{subsec:hypoercube permutation} 
Before we proceed to the description of the general achievable scheme, we introduce two definitions of special permutations on a given set of points, i.e., the \tit{hypercube permutation} and \tit{circular hypercube permutation}, which are essential to 
the description of the general delivery phase. 
\begin{defn}
\label{defn_hcb_perm}
\textbf{(Hypercube Permutation)} Given a set of $D\times t$ points, denoted by $\mathcal{Q}$, i.e., $|\cq|=Dt$, we label each of these points by a unique number $u_{i,j}\in[Dt]$, where $i\in[t],j\in[D]$. Assume that these points are partitioned into $t$ disjoint groups, which we refer to as \textit{dimensions}. Each dimension consists of $D$ points, denoted by $\mathcal{U}_i=\left\{u_{i,j}:\lfloor\frac{u_{i,j}}{D}\rfloor=i,j=0,1,\cdots,D-1\right\}$, $i\in[t]$. Define a \textit{hypercube permutation} of the set $\mathcal{Q}$, denoted by $\mathbf{\pi}^{\rm HCB}=[\,\pi(0)\;\pi(1)\;\cdots\;\pi(Dt-1)\,]$, as such a permutation of the $Dt$ points that satisfies the following condition: For any set of points $\mathcal{U}_i,i\in[t]$, the positions in the permutation (denoted by $pos(\cdot)$, meaning that $pos(u)=i$ if $\pi(i)=u$) of any two of them, $u_{i,j_1}$ and $u_{i,j_2}\,(j_1\neq j_2)$, should satisfy $|pos(u_{i,j_1})-pos(u_{i,j_2})|=kt,1\leq k \leq D-1,k\in \mathbb{Z^+}$ and $j_1,j_2\in[D]$.
\hfill $\lozenge$ 
\end{defn}
\begin{defn}
\label{defn_hcb_perm_circ}
\textbf{(Circular Hypercube Permutation)} {A \emph{circular permutation} of a set $\mathcal{Q}$ is a way of arranging the elements of $\mathcal{Q}$ such that these arrangements are invariant of circular shifts. Denote the set of circular permutations of $\mathcal{Q}$ as $\Pi_{\mathcal{Q}}^{\rm circ}$. For example, if $\mathcal{Q}=\{1,2,3\}$, then $\Pi_{\mathcal{Q}}^{\rm circ}=\{[1\; 2\; 3\;],[1\; 3\; 2\;]\}$.} A \textit{circular hypercube permutation} of a set $\mathcal{Q}$ is a way of arranging the elements of $\mathcal{Q}$ around a fixed table, and meanwhile, the corresponding arrangement should be a hypercube permutation.
\hfill $\lozenge$
\end{defn}

We illustrate the concept of hypercube permutation and circular hypercube permutation via the following example. 
\begin{example}
For $\mathcal{Q}=\{0,1,2,3\}$ with $t=2$ dimensions and $D=2$ points in each dimension, i.e., $\mathcal{U}_0=\{0,1\},\,\mathcal{U}_1=\{2,3\}$, we have 
\begin{eqnarray*}
\Pi^{\rm HCB}_{\mathcal{Q}}=\Big\{[\,0\;2\;1\;3\,],[\,0\;3\;1\;2\,],[\,1\;2\;0\;3\,],[\,1\;3\;0\;2\,],\\
\qquad[\,2\;1\;3\;0\,],[\,2\;0\;3\;1\,],[\,3\;1\;2\;0\,],[\,3\;0\;2\;1\,]\Big\}.
\end{eqnarray*}
It is clear that, for any two points within one dimension, $0,1\in\mathcal{U}_0$ or $2,3\in\mathcal{U}_1$, we have $|pos(0)-pos(1)|=|pos(2)-pos(3)|=2$, which satisfies the condition $|pos(u_{i,j_1})-pos(u_{i,j_2})|=t$ (note that $k=1$). Furthermore, we have $\Pi^{\rm HCB,circ}_{\mathcal{Q}}=\big\{[\;0\;2\;1\;3\;], [\;0\;3\;1\;2\;]\big\}$. 
\hfill $\triangle$
\end{example}

\begin{lemma}
\label{lemma_1}
\tit{For a set of points (users) $\mathcal{Q}$ of dimension $t$ and $D$ points (users) in each dimension, denote the set of all hypercube permutations as $\Pi_{\mathcal{Q}}^{\rm HCB}$, then $
\left|\Pi_{\mathcal{Q}}^{\rm HCB}\right|=(D!)^t(t)!
$. The set of circular hypercube permutations of $\mc{Q}$, denoted by $\Pi_{\mathcal{Q}}^{\rm HCB, circ}$, has size $
\left|\Pi_{\mathcal{Q}}^{\rm HCB, circ}\right|=\frac{(D!)^t(t-1)!}{D}
$}. \hfill $\square$
\end{lemma}
\begin{IEEEproof}
See Appendix \ref{appendix_1}.
\end{IEEEproof}

\subsection{General Achievable Scheme}
\label{subsec:General Achievable Scheme}
In this section, we present the general achievable scheme which is formally described in Algorithm \ref{algorithm:1}. Recall that $t_T= \frac{K_TM_T}{N}$ and $t_R= \frac{K_RM_R}{N}$, and we assume $t_T, t_R \in \mathbb{Z}^+,\tron\le \kr$. In this paper, we focus on the case $\delta\triangleq\frac{t_T}{t_R}\in \mathbb{Z}^+$, implying that {$\ttt\ge1$}. 
\begin{algorithm*}
\setstretch{1.00}   
  \caption{General Hypercube-based Achievable Scheme}\label{algorithm:1}
  \begin{algorithmic}[1]
  \item[\quad$\,\,$ \textit{Prefetching Phase:}] 
  \For {$i=0,1,\cdots,K_T-1$}
  \State Group Tx$_i$ into the transmitter dimension $\mc{U}^{\rm T}_{j}$, where $j=\lfloor\frac{i}{D_T}\rfloor$. 
  \EndFor
  \For {$i=0,1,\cdots,K_R-1$}
  \State Group Rx$_i$ into the receiver label set $\mathcal{U}^{\rm R}_{j}$, where $j=\lfloor\frac{i}{D_R}\rfloor$. 
  \EndFor
  \For {$n=0,1,\cdots,N-1$}
  \State Split $\mathcal{W}_n$ into $(\frac{N}{M_T})^{t_T}(\frac{N}{M_R})^{t_R}$ disjoint equal-size subfiles: $$\mathcal{W}_n=\left\{\mathcal{W}_{n,\mathcal{T},\mathcal{R}}\right\}_{\begin{array}{c}\mathcal{T}\in \mathcal{U}^{\rm T}_0\bigotimes\mathcal{U}^{\rm T}_1\bigotimes\cdots\bigotimes\mathcal{U}^{\rm T}_{t_T-1}\\
\mathcal{R}\in \mathcal{U}^{\rm R}_0\bigotimes\mathcal{U}^{\rm R}_1\bigotimes\cdots\bigotimes\mathcal{U}^{\rm R}_{t_R-1}\end{array}}$$.
  \EndFor 
  \For {$i=0,1,\cdots,K_T-1$}
  \State Tx$_i$ caches $\left\{\mathcal{W}_{n,\mathcal{T},\mathcal{R}}:i\in \mathcal{T}\right\}$ for all $n\in[N]$. 
  \EndFor 
  \For {$j=0,1,\cdots,K_R-1$}
  \State Rx$_j$ caches $\left\{\mathcal{W}_{n,\mathcal{T},\mathcal{R}}:j\in \mathcal{R}\right\}$ for all $n\in[N]$. 
  \EndFor  
  \item[]
  \item[\quad$\,\,$ \textit{Delivery Phase:}]   
  \For {$j=0,1,\cdots,K_R-1$}
  \For {$\mathcal{T}\in\mathcal{U}^{\rm T}_0\bigotimes\mathcal{U}^{\rm T}_1\bigotimes\cdots\bigotimes \mathcal{U}^{\rm T}_{t_T-1}$}
  \For {$\mathcal{R}\in\mathcal{U}^{\rm R}_0\bigotimes\mathcal{U}^{\rm R}_1\bigotimes\cdots\bigotimes\mathcal{U}^{\rm R}_{\lfloor\frac{j}{D_R}\rfloor}\setminus \{j\} \bigotimes\cdots \bigotimes\mathcal{U}^{\rm R}_{t_R-1}$} 
  \State {Split the subfile $\mathcal{W}_{d_j,\mathcal{T},\mathcal{R}}$ into $\binom{D_R-2}{\delta-1}{\binom{D_R-1}{\delta}}^{t_R-1}\frac{(\delta!)^{t_R}}{\delta}(t_R-1)!$ disjoint packets 
  
of eqaul-sizes:  
$$
\left\{\mathcal{W}_{d_j,\mathcal{T},\dot{\pi},\ddot{\pi}}\right\}_{\subalign {&\dot{\pi}=\pi[1:t_R]\\
&\ddot{\pi}=\pi[t_R+1:t_T+t_R-1]\\
&\pi\in\Pi^{\rm HCB}_{\mathcal{Q}^{\rm U}},\,\pi(0)=j,\,\pi(t_R)=r_{\lfloor\frac{j}{D_R}\rfloor}\\
&\left\{\pi(1),\pi(2),\cdots,\pi(t_R-1)\right\}=\mathcal{R}\setminus\{r_{\lfloor\frac{j}{D_R}\rfloor}\}\\  }}
$$
\qquad where $\mathcal{Q}\in\Gamma_{\mathcal{U}^{\rm R}_0,\delta+1}\bigotimes\cdots\bigotimes\Gamma_{\mathcal{U}^{\rm R}_{\lfloor\frac{j}{D_R}\rfloor},\delta+1}\bigotimes\cdots\bigotimes\Gamma_{\mathcal{U}^{\rm R}_{t_R-1},\delta+1}$. }
  \EndFor
  \EndFor  
  \EndFor 
  \For {$\mathcal{T}\in\mathcal{U}^{\rm T}_0\bigotimes\mathcal{U}^{\rm T}_1\bigotimes\cdots\bigotimes \mathcal{U}^{\rm T}_{t_T-1}$} 
  \For {$\mathcal{R}\in\Gamma_{\mathcal{U}^{\rm R}_0,\delta+1}\bigotimes\Gamma_{\mathcal{U}^{\rm R}_1,\delta+1}\bigotimes\cdots\bigotimes\Gamma_{\mathcal{U}^{\rm R}_{t_R-1},\delta+1}$}
  \For {$\pi\in\Pi_{\mathcal{R}^{\rm U}}^{\rm HCB,circ}$}
  \State Each transmitter sends a linear combination (Lemma \ref{lemma:existence of linear coefficients}) of the coded packets:
  $$\qquad \qquad S_i=\mathcal{L}_{i,\mathcal{T},\pi}\left(\left\{  \hat{W}_{{d_{\pi(\ell)}},\mathcal{T}(\ell),\pi[\ell+1:\ell+t_R],\pi[\ell+t_R+1:\ell+t_R+t_T-1]}:\ell\in[t_T+t_R],i\in\mathcal{T}(\ell) \right\}\right)$$
  \EndFor
  \EndFor
  \EndFor   
  \end{algorithmic}
\end{algorithm*}

 
The corresponding prefetching and delivery phases are described as follows.
\subsubsection{Prefetching Phase}
The hypercube cache placement is employed at both the transmitters' and receivers' sides in the prefetching phase. Refer to Section \ref{subsubsec:Hypercube cache placement design for cache-aided interference networks} for detailed descriptions.

\subsubsection{Delivery Phase}
In the delivery phase, the receivers' demand vector $\mathbf{d}=[d_0,d_1,\cdots,d_{K_R-1}]$ is revealed, i.e., each receiver Rx$_j,j \in [K_R]$ requests a file $\mathcal{W}_{d_j}$. Since some subfiles of the requested file have already been cached by the receiver in the prefetching phase, the transmitters only need to send those subfiles which have not been cached by Rx$_j$, i.e., $\{\mathcal{W}_{d_j,\mathcal{T}},\forall \mathcal{T},\forall\mathcal{R}:j\notin\mathcal{R}\}$.  

Following a similar methodology of \cite{Naderializadeh2017interference}, we need to further split the set of subfiles to be delivered to the receivers into 
packets so that they can be scheduled in subsets of size $t_T+t_R$ and delivered to the receivers simultaneously without interference. 
In particular, for any 
packet in the subset of 
$t_T+t_R$ packets, it is requested by one particular receiver and can be cancelled by another $t_R$ receivers by utilizing their cached packets. Also, the transmitters can collaborate to zero-force the the interference to another $t_T-1$ unintended receivers. We describe how to do such a further splitting based on the hypercube cache placement in the following.

For any $j\in[\kr]$, $\mathcal{T}=\{(\tau_0,\tau_1,\cdots,\tau_{\ttt -1})\}$ with $(\tau_0,\tau_1,\cdots,\tau_{\ttt -1})\in\mathcal{U}^{\rm T}_0\times\mathcal{U}^{\rm T}_1\times\cdots\times \mathcal{U}^{\rm T}_{t_T-1}$, and $\mathcal{R}=\{(r_0,r_1,\cdots,r_{\ttt -1})\}$ with $(r_0,r_1,\cdots,r_{\ttt -1})\in\mathcal{U}^{\rm R}_0\times\mathcal{U}^{\rm R}_1\times\cdots\times\mathcal{U}^{\rm R}_{\lfloor\frac{j}{D_R}\rfloor}\setminus \{j\} \times\cdots \times\mathcal{U}^{\rm R}_{t_R-1}$ (note that $|\mathcal{T}|=t_T$ and $|\mathcal{R}|=t_R$), we split $\mathcal{W}_{d_j,\mathcal{T},\mathcal{R}}$ into $\binom{D_R-2}{\delta-1}{\binom{D_R-1}{\delta}}^{t_R-1}\frac{(\delta!)^{t_R}}{\delta}(t_R-1)!$ disjoint packets of equal-sizes,  denoted by 
\begin{equation}
\left\{\mathcal{W}_{d_j,\mathcal{T},\dot{\pi},\ddot{\pi}}\right\}_{\subalign {&\dot{\pi}=\pi[1:t_R]\\
&\ddot{\pi}=\pi[t_R+1:t_T+t_R-1]\\
&\pi\in\Pi^{\rm HCB}_{\mathcal{Q}^{\rm U}},\,\pi(0)=j,\,\pi(t_R)=r_{\lfloor\frac{j}{D_R}\rfloor}\\
&\left\{\pi(1),\pi(2),\cdots,\pi(t_R-1)\right\}=\mathcal{R}\setminus\{r_{\lfloor\frac{j}{D_R}\rfloor}\}\\  }},
\end{equation}
where $\mathcal{Q}\in\Gamma_{\mathcal{U}^{\rm R}_0,\delta+1}\bigotimes\cdots\bigotimes\Gamma_{\mathcal{U}^{\rm R}_{\lfloor\frac{j}{D_R}\rfloor},\delta+1}\bigotimes\cdots\bigotimes\Gamma_{\mathcal{U}^{\rm R}_{t_R-1},\delta+1}$ and the notations are defined as follows. For a set $\mathcal{S}$, $\Gamma_{\mathcal{S},s}$ is defined as a set whose elements are all subsets of $\mathcal{S}$ of size $s$, i.e.,
$\Gamma_{\mathcal{S},s}=\left\{\mathcal{A}:\mathcal{A}\subseteq \mathcal{S},|\mathcal{A}|=s\right\},\,s=1,2,\cdots,|\mathcal{S}|$. For example, for $\mathcal{S}=\left\{0,1,2\right\}$, we have $\Gamma_{\mathcal{S},2}=\left\{\{0,1\},\{1,2\},\{0,2\}\right\}$. For a set $\mathcal{Q}$ whose elements are sets, $\mathcal{Q}^{\rm U}$ denotes the union of the elements in $\mathcal{Q}$. For example, if $\mathcal{Q}=\left\{\{0,1\},\{2,3\}\right\}$, we have $\mathcal{Q}^{\rm U}=\{0,1\}\cup \{2,3\}=\{0,1,2,3\}$. Moreover, for a set $\mathcal{S}$, and a hypercube permutation $\pi\in \Pi^{\rm HCB}_{\mathcal{S}}$ and two integers $i,j$, where $i\leq j$, $\pi[i:j]$ is defined as $\pi[i:j]=[\pi(i\oplus_{|\mathcal{S}|}0)\;\pi(i\oplus_{|\mathcal{S}|}1),\;\cdots,\;\pi(i\oplus_{|\mathcal{S}|}(j-i))]$, in which for two integers $m,n$, $m\oplus_{|\mathcal{S}|}n$ is defined as
\be 
m\oplus_{|\mathcal{S}|}n=1+(m+n-1 \mod |\mathcal{S}|).
\ee

After such a further splitting, for a specific set of $t_T+t_R$ receivers and a corresponding hypercube permutation $\pi$, the packet $\mathcal{W}_{d_j,\mc{T},\dot{\pi},\ddot{\pi}}$, which is desired by Rx$_j$, can be cancelled at receivers in $\dot{\pi}$ by utilizing their individual cached contents and can be zero-forced at receivers in $\ddot{\pi}$ through the collaboration of some transmitters. Lemma \ref{lemma:grouping} shows how this further splitting is done. For a set $\mathcal{T}=\{\tau_0,\tau_1,\cdots,\tau_{t_T-1}\}$ whose elements are from the $t_T$ different transmitter dimensions, i.e., $\tau_i\in\mathcal{U}^{\rm T}_{i}, i\in[\ttt]$, we define the corresponding sets $\mathcal{T}(\ell)\triangleq\{\tau_0^{(\ell)},\tau_1^{(\ell)},\cdots,\tau_{t_T-1}^{(\ell)}\},\ell \in [\ttt +\ttr]$, where $\mathcal{T}(0)=\mathcal{T}$, i.e., $\tau_i^{(0)}=\tau_i,\forall i\in[t_T]$ and

$\bullet$ When $1\leq\ell\leq t_T$,
\begin{equation}
\tau_i^{(\ell)}=\left\{
\begin{array}{ll}
\tau_i ^{(0)}+1\mod D_T& 0\leq i\leq \ell-1, \\
\tau_i^{(0)} &\ell \leq i\leq t_T-1. \\
\end{array}\right. 
\end{equation}

$\bullet$ When $t_T+1\leq\ell\leq t_T+t_R-1$, 
\begin{equation}
\tau_i^{(\ell)}=\left\{
\begin{array}{ll}
\tau_i^{(0)} & 0\leq i\leq \ell-t_T-1 ,\\
\tau_i ^{(0)}+1\mod D_T &\ell-t_T \leq i\leq t_T-1.\\
\end{array} \right.
\end{equation}

\begin{lemma}
\label{lemma:grouping}\textit{Based on the hypercube cache placement, for any receivers' demand vector $\mathbf{d}$, the set of packets 
needed to be sent to the receivers can be grouped into disjoint subsets of size $t_T+t_R$ as}
\begin{multline}
\bigcup_{\substack{\mathcal{T}\in\mathcal{U}^{\rm T}_0\bigotimes\mathcal{U}^{\rm T}_1\bigotimes\cdots\bigotimes \mathcal{U}^{\rm T}_{t_T-1}\\
\mathcal{R}\in\Gamma_{\mathcal{U}^{\rm R}_0,\delta+1}\bigotimes\Gamma_{\mathcal{U}^{\rm R}_1,\delta+1}\bigotimes\cdots\bigotimes\Gamma_{\mathcal{U}^{\rm R}_{t_R-1},\delta+1}\\
\pi\in\Pi_{\mathcal{R}^{\rm U}}^{\rm HCB,circ}}} 
\times \left\{\mathcal{W}_{{d_{\pi(\ell)}},\mathcal{T}(\ell),\pi[\ell+1:\ell+t_R],\pi[\ell+t_R+1:\ell+t_R+t_T-1]}:
\ell\in[\ttt+\ttr]\right\},\label{Eqn:grouping}
\end{multline}
\end{lemma}

\begin{IEEEproof}
See Appendix \ref{sec: proof of lemma 2}.  
\end{IEEEproof}

Given the grouping method of the packets in Lemma 2, we will have $D_T^{t_T}\binom{D_R}{\delta+1}^{t_R}\frac{[(\delta+1)!]^{t_R}(t_R-1)!}{\delta+1}$ (using Lemma \ref{lemma_1}) steps of communications. More specifically, the term $\dt^{\ttt}$ corresponds to the number of possible choices of $\mc{T}$, $\binom{\dr}{\delta +1}^{\ttr}$ corresponds to the number of choices of $\mc{R}$. We also have $|\Pi_{\mathcal{R}^{\rm U}}^{\rm HCB,circ}|=\frac{[(\delta+1)!]^{t_R}(t_R-1)!}{\delta+1}$ which is a direct result of Lemma \ref{lemma_1}, i.e., the number of different hypercube permutations of the set $\mc{R}^{\rm U}$ partitioned into $t=\ttr$ dimensions and $D=\delta +1$ points in each dimension. In each of these communication steps, specific sets $\mathcal{T}$ and $\mathcal{R}$ and a hypercube permutation are fixed, and each transmitter Tx$_i,i\in \mathcal{T}(\ell)$ transmits a linear combination of the coded packets, i.e.,
\begin{equation}\label{linear combo}
S_i=\mathcal{L}_{i,\mathcal{T},\pi}\left(\left\{\hat{W}_{d_{\pi(\ell)},\mathcal{T}(\ell),\pi[\ell+1:\ell+t_R],\pi[\ell+t_R+1:\ell+t_R+t_T-1]}:
\ell\in[t_T+t_R],i\in\mathcal{T}(\ell)\right\}\right), 
\end{equation}
in which for any packet $\mathcal{W}_{d_j,\mathcal{T},\dot{\pi},\ddot{\pi}}$, $\hat{W}_{d_j,\mathcal{T},\dot{\pi},\ddot{\pi}} $ denotes its coded version, and $\mathcal{L}_{i,\mathcal{T},\pi}(.)$ represents the linear combination that Tx$_i$ chooses to transmit set of packets in (\ref{linear combo}).

The following lemma shows the existence of the linear combination coefficients.

\begin{lemma}\label{lemma:existence of linear coefficients} \textit{For any subset of $t_T$ transmitters $\mathcal{T}\in\mathcal{U}^{\rm T}_0\bigotimes\mathcal{U}^{\rm T}_1\bigotimes\cdots\bigotimes \mathcal{U}^{\rm T}_{t_T-1}$, any set of $t_T+t_R$ receivers $\mathcal{R}^{\rm U}$ for which $\mathcal{R}\in\Gamma_{\mathcal{U}^{\rm R}_0,\delta+1}\bigotimes\Gamma_{\mathcal{U}^{\rm R}_1,\delta+1}\bigotimes\cdots\bigotimes\Gamma_{\mathcal{U}^{\rm R}_{t_R-1},\delta+1}$, and any circular hypercube permutation $\pi\in\Pi_{\mathcal{R}^{\rm U}}^{\rm HCB,circ}$, there exists a choice of the linear combinations $\{\mathcal{L}_{i,\mathcal{T},\pi}(.)\}_{i=1}^{K_T}$ in (\ref{linear combo}) such that the set of $t_T+t_R$ packets in 
\begin{eqnarray}
\left\{\mathcal{W}_{{d_{\pi(\ell)}},\mathcal{T}(\ell),\pi[\ell+1:\ell+t_R],\pi[\ell+t_R+1:\ell+t_R+t_T-1]}:\ell\in[\ttt+\ttr]\right\}
\end{eqnarray} can be delivered simultaneously without interference by the transmitters in $\bigcup_{\ell\in[t_T+t_R]}\mathcal{T}(\ell)$ to the receivers in $\mathcal{R}^{\rm U}$.}
\end{lemma}
\begin{IEEEproof}
The proof of Lemma 3 follows exactly the same steps given in \cite{Naderializadeh2017interference}. To show the existence of such linear combinations, we require the linear coefficients to be designed such that for any receiver in $\mathcal{R}^{\rm U}$, its desired packets must be received with non-zero coefficients, and the undesired subfiles which can not be cancelled by utilizing its cached content, must be zero-forced. Then we can show the existence of such linear combinations simply by observing the fact that the number of variables (coefficients) equals the number of equations (received signal requirements). The details of proof are omitted here.  
\end{IEEEproof}

\subsection{Subpacketization Complexity Analysis}
\label{complexity analysis}
In this section , we provide a comprehensive performance comparison between the proposed hypercube-based 
based scheme and the NMA scheme. 

In the hypercube-based scheme, each file in the library is split into $(\frac{N}{M_T})^{t_T}(\frac{N}{M_R})^{t_R}$ subfiles while in
the NMA scheme each file is split into $\binom{K_T}{t_T}\binom{K_R}{t_R}$ subfiles. 
In the delivery phase, to implement interference cancellation, each requested subfile 
is further split into 
\be \Delta_{\rm HCB}(\kt,\mt,\kr,\mr,N)\triangleq\binom{D_R-2}{\delta-1}{\binom{D_R-1}{\delta}}^{t_R-1}\frac{(\delta!)^{t_R}}{\delta}(t_R-1)!\ee packets in the proposed hypercube-based scheme and \be\Delta_{\rm NMA}(\kt,\mt,\kr,\mr,N)\triangleq \binom{K_R-t_R-1}{t_T-1}(t_T-1)!t_R!\ee packets in the NMA scheme.  
{To measure the subpacketization complexity, we count the total number of packets that a specific file needs to be split into which equals the number of subfiles per file times the number of packets per subfile. When counting the number of subfiles per file, both the pre-stored and requested subfiles by any receiver should be included 
since the total number of packets per file should reflect the size ($L/F$ bits) of the smallest units (i.e., packets) that a file is split into. Therefore, the total number of packets for a specific requested file required for these two schemes are 
\begin{eqnarray}
F_{\rm HCB}(\kt,\mt,\kr,\mr,N)&=& {D_T}^{\ttt}{D_R}^{\ttr}\Delta_{\rm HCB}(\kt,\mt,\kr,\mr,N), \\
F_{\rm NMA}(\kt,\mt,\kr,\mr,N) &=& \binom{K_T}{t_T}\binom{K_R}{t_R}\Delta_{\rm NMA}(\kt,\mt,\kr,\mr,N).
\end{eqnarray}}Since the comparison of subpacketization levels is always done under the same set of system parameters, we ignore the these parameters in the expressions of $\Delta_{\rm HCB},\Delta_{\rm NMA},F_{\rm HCB}$ and $F_{\rm NMA}$ for brevity. 
\begin{figure}
\begin{center}
\subfigure[Illustration of $G(d,t,\delta)$ as a function of $t$ when $\delta =1$.]{
\centering \includegraphics[width=0.4\textwidth]{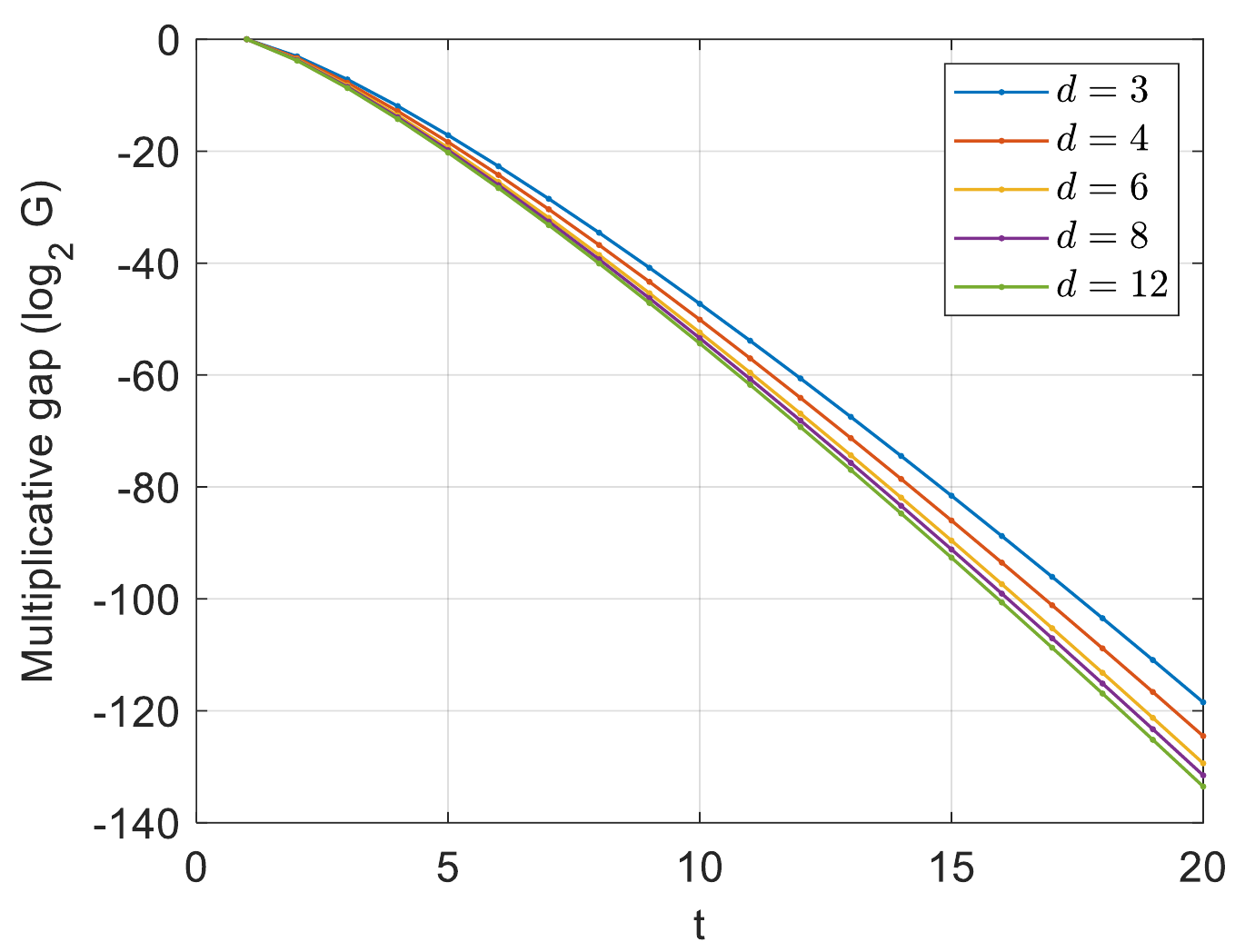}} 
\subfigure[Illustration of $G(d,t,\delta)$ as a function of $t$ when $\delta =2$.]{
\centering \includegraphics[width=0.4\textwidth]{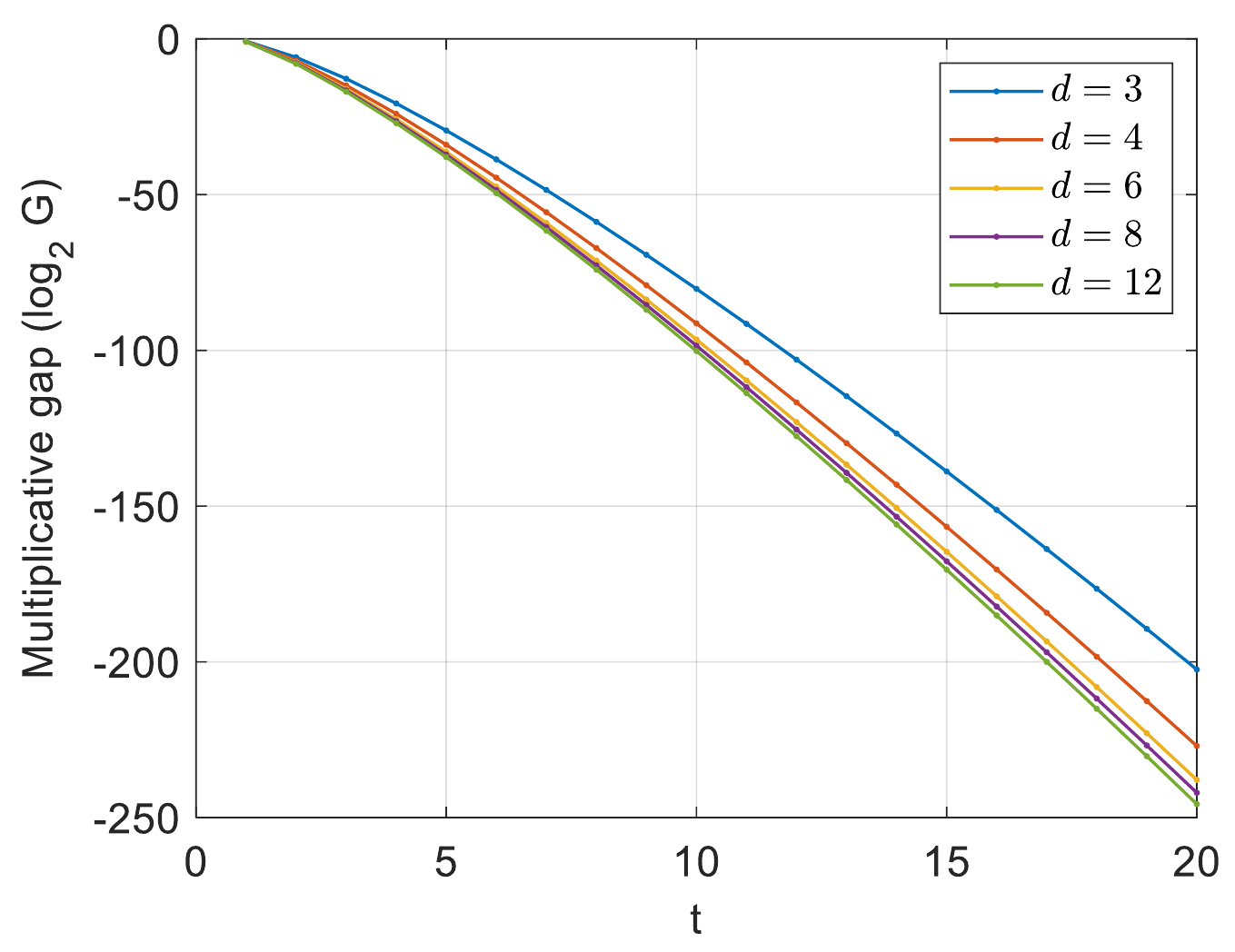}} 
\subfigure[No. of subfiles per-file gap as a function of $t$ when $\delta =1$.]{
\centering \includegraphics[width=0.4\textwidth]{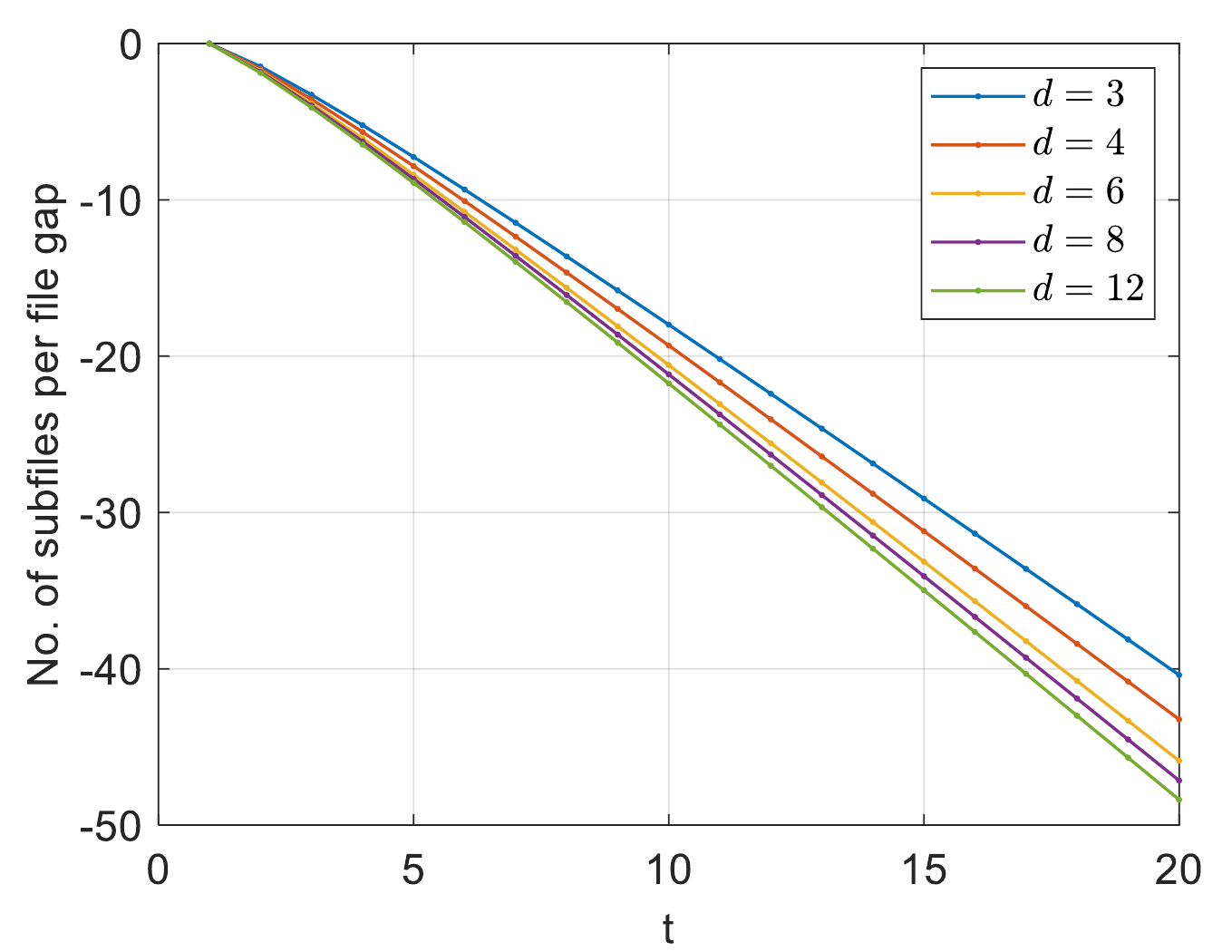}} 
\subfigure[No. of packets per-subfile gap as a function of $t$ when $\delta =1$.]{
\centering \includegraphics[width=0.4\textwidth]{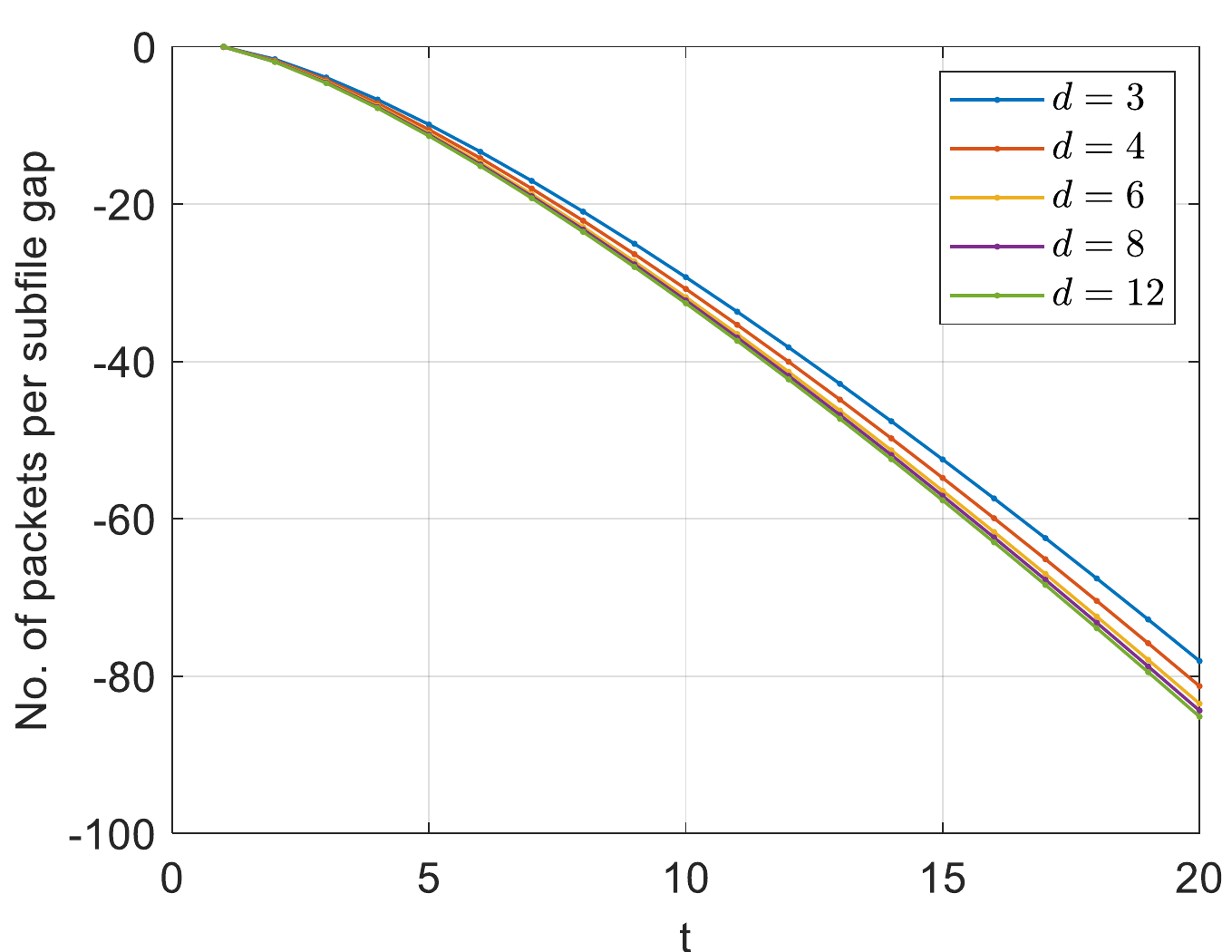}} 
\end{center}
\caption{The multiplicative gap $G$ between the hypercube scheme and the NMA scheme. The comparison is down under the setting $t_T=\delta t_R=\delta t$, $N/M_T=N/M_R=d$, which implies $K_T=\delta K_R=\delta dt$. It can be seen that: (a) $\delta =1$. For a fixed $d$, $G$ decreases (exponentially) quickly as $t$ increases and approaches zero as $t$ goes to infinity, and (b) $\delta =2$. In this case, the number of transmitters increases and $G$ decreases faster. (c) $\delta =1$. Comparison of the No. of subfiles per file of the hypercube design to the NMA scheme (logarithmic scale). (d) $\delta =1$. Comparison of the No. of packets per subfile of the two schemes (logarithmic scale). }
\label{figure:4}
\end{figure} 
To compare the subpacketization level between our scheme and the NMA scheme, we define the multiplicative gap of the subpacketization levels between these two schemes as follows.

\begin{defn} \tbf{(Multiplicative Gap of Subpacketization Levels)} For the system parameters $\kt,\mt,\kr,\mr$ and $N$, the \emph{multiplicative gap $G$} of the subpacketization levels between the hypercube-based scheme and the NMA scheme, is defined as
\be
G(\kt,\mt,\kr,\mr,N)\triangleq \frac{F_{\rm HCB}(\kt,\mt,\kr,\mr,N)}{F_{\rm NMA}(\kt,\mt,\kr,\mr,N)}.
\ee
For ease of notation, we ignore the parameters and simply denote $G={F_{\rm HCB}}/{F_{\rm NMA}}$. \hfill $\lozenge$   
\end{defn}

We next show that for any system parameters, the hypercube scheme has a strictly less subpacketization level than that of the NMA scheme. Moreover, we show that there is an order gain compared to the NMA scheme when $t\to \infty$  if $d$ and $\delta$ are fixed. 
 {
\begin{theorem}
\label{theorem:2} For any system parameters $K_T,K_R,M_T,M_R$ and $N$ satisfying $t_T=\frac{K_TM_T}{N}\in \mathbb{Z}^+,t_R=\frac{K_RM_R}{N}\in \mathbb{Z}^+,D_T=\frac{K_T}{t_T}\in\mathbb{Z}^+,D_R=\frac{K_R}{t_R}\in\mathbb{Z}^+$ and $\delta\triangleq\frac{t_T}{t_R}\in\mathbb{Z}^+,D_{R}\ge \delta +1$, the  multiplicative gap $G$ is strictly less than $1$. Moreover,  
\begin{equation}
\label{eq: thm 2, order gain t}
G(d,t,\delta) \le  C_0\left(\frac{C_1}{t} \right)^t\frac{1}   {t^{(\delta -1) t-1}},
\end{equation} 
 where $C_0=\frac{(d-2)!e^6}{(d-\delta -1)!} \left(\sqrt{\frac{d-1}{\delta}}\frac{(d-1)!}{(d-\delta -1)!}\left( \frac{2\pi}{d-1} \right)^{3/2}\right)^{-1}$ and 
 \begin{equation}
 C_1 = \frac{(d-1)!}{(d-\delta -1 )!}\left( \frac{e}{d-1} \right)^{\delta}\left(  \frac{d}{d-1} \right)^{-(\delta +1)(d-1)}\left( \frac{d-1}{d-\delta-1}\right)^{-(d-\delta-1)}.  \end{equation}
 \hfill $\square$ 
\end{theorem}}
\begin{IEEEproof}
 See Appendix \ref{sec: proof of Theorem 2}.
\end{IEEEproof}

{Theorem 2 shows that the proposed hypercube-based scheme strictly outperforms the NMA scheme in terms of subpacketizaiton while achieving the same one-shot liner sum-DoF. In fact, the proposed scheme requires not only a smaller number of subfiles per file but also a smaller number of packets per subfile than the NMA scheme, demonstrating the advantage of the hypercube-based design. From (\ref{eq: thm 2, order gain t}) we see  that if $t\ge C_1$, $G(d,t,\delta)\le C_0/t^{(\delta-1)t-1}$. Therefore, for fixed $d$ and $\delta$, we have the scaling $G(d,t,\delta)=O(1/t^{(\delta-1)t-1})$ as $t\to \infty$, implying that there is an order gain in subpacketization of the hypercube-based scheme compared to the NMA scheme.  Fig.~\ref{figure:4} shows the multiplicative gain $G(d,t,\delta)$ under logarithmic scale for the case when $\delta\triangleq {t_T}/{t_R}=1,2$ under the setting $t_T=\delta t_R =\delta t, D_R=D_R=d$. It can be seen that the gap decreases exponentially as $t$ increases and goes to zero as $t$ goes to infinity (see Fig. \ref{figure:4}(a), (b)), which demonstrates an order gain in subpacketization reduction of the proposed scheme compared to the NMA scheme. Moreover, from Fig.~\ref{figure:4} (c),(d), it can be seen that the proposed scheme also requires exponentially smaller number of subfiles per file and packets per subfile.} 

\section{Discussion}
In this section, we will first provide two possible extensions of the proposed scheme, which are cache-aided Device-to-Device (D2D) interference networks and wireless coded distributed computing networks.
Second, we will discuss the connection to and differences from some related existing works.

\subsection{Extension to cache-aided D2D Interference Networks and Wireless Distributed Computing Systems}
In the settings of a typical cache-aided D2D interference 
networks, all the nodes (or devices) are expected to have homogeneous cache memory sizes. 
The proposed hypercube-based scheme can be directly extended to such D2D interference networks to achieve an order-optimal one-shot linear sum-DoF while maintaining the promised 
subpacketization levels compared to the direct translation of the NMA scheme. There are multiple approaches to apply the hypercube-based approach to cache-aided D2D interference networks. In the following, we will illustrate one example of such applications. 
We consider a D2D interference network with a library of $N$ files and $K$ nodes, each equipped with a cache memory of size $M$ files. We assume $K$ is even and $t = KM/N \leq K/2$. We partition the network into two groups with equal number of devices, i.e., each group has $K/2$ devices. Let $t' = \frac{KM}{2N} \in \mathbb{Z}^+$. In the prefetching phase, in each group, we perform the hypercube cache placement such that the two groups have identical cache placement. The delivery phase has two steps, in the first step, one group of nodes will perform as transmitters and the other group will perform as receivers. Note that since $\kt = \kr = K/2$, the proposed delivery scheme based on the hypercube cache placement can be directly used. The achievable sum-DoF is $t = \ttt + \ttr = KM/N$. In the first phase, the requests from one group of receivers can be served. In the second step, we exchange the groups of transmitters and receivers such that the other group can be served with the same achievable sum-DoF. Therefore, the total achievable sum-DoF is given by $t= KM/N$. 

Moreover, due to the similarity between the cache-enabled D2D interference network and the Coded Distributed Computing (CDC, \cite{li2018fundamental}), the hypercube cache placement can be directly applied to the wireless CDC interference networks. 
From the wireless D2D caching network example, it can be seen that the proposed hypercube-based scheme can be applied in a more practical half-duplex transmission settings. For example, 
the hypercube cache placement scheme 
can be employed in the file assignment phase in the CDC networks. Then we use the same delivery scheme as in the wireless D2D caching networks to achieve an order optimal communication-computation trade-off. 

{\subsection{Comparison with Existing Works }
In this section we discuss the connection to and the differences from the most related works \cite{lampiris2018adding} and \cite{salehi2020low} and thus  highlight the uniqueness of the hypercube-based design.

The work of \cite{lampiris2018adding} shows that by adding multiple ($L$) transmit antennas the supacketization level of coded caching can be reduced approximately to its $L$-th root compared to the shared-link coded caching scheme. It turns out that this scheme can be extended to the cache-aided $\kt \times \kr$ interference networks to achieve the same sum-DoF as the hypercube-based scheme proposed. However, due to the use of user/receiver grouping, the scheme of \cite{lampiris2018adding} suffers sum-DoF loss (i.e., can not achieve $t_T+t_R$) when either $K_R/t_T$ or $t_R/t_T$ is not an integer, which means that it can not achieve sum-DoF $t_T+t_R$ when $\delta =t_T/t_R>1$, putting a major limitation to its applicability. Moreover,  \cite{salehi2020low}   considered a similar setting as \cite{lampiris2018adding} but a totally different cyclic cache placement based on PDA was proposed to achieve the sum-DoF $K\gamma +L$ with a quadratic (w.r.t. $K$) subpacketization. However, the proposed scheme works only when $K\gamma \le L$. The differences of our work from these two works are summarized as follows.

\emph{1) Different design methodologies and parameter regimes}. The hypercube-based scheme applies the hypercube cache
placement with a nice geometric interpretation and does not rely on receiver grouping which requires that $t_R\ge t_T$)
puts a strong limitation in applying to interference networks. In contrast, our work
primarily focused on the case $\delta \ge 1$, although by using a memory sharing alike method,
the hypercube-based design can be extended to the cases when $\delta $ is not an integer or $\delta<1$ 
without sum-DoF loss (Note that the scheme of  \cite{salehi2020low} does not work when $\delta <1$). In this regime, At the point $\delta=1$, the scheme of \cite{lampiris2018adding} achieves a lower subpacketization level than the hypercube-based scheme which can be shown as follows. Assume $M_T\in\mathbb{Z}^+$ (otherwise the transmitter side cache placement in \cite{lampiris2018adding} does not work), then \cite{lampiris2018adding} requires subpacketization $\binom{K_T/t_T}{t_R/t_T}=K_T/t_T=D_T$ while the hypercube-based scheme has $F_{\rm HCB}= D_T^{t_T}D_R^{t_R}\Delta_{\rm HCB}$ which is larger than $D_T$.  

\emph{2) Symmetry in Cache Placement. } Different from the hypercube-based design, 
\cite{lampiris2018adding} employs asymmetric cache placement methods at the transmitter and receiver sides. One potential drawback is that the scheme of \cite{lampiris2018adding} can not be directly applied to cache-aided Device-to-Device (D2D) interference networks and the wireless CDC systems where each user needs to be both transmitter and receivers in order to fulfill the file requests of all users. However, due to the symmetric cache placements at both the transmitter and receiver sides, the hypercube-based scheme extends naturally to such networks and incurs no extra cost when the users switch their roles from transmitters to receivers or vice versa.

}

\if{0}
The difference between the hypercube-based scheme and \cite{lampiris2018adding} is apparent: the subpacketization reduction gain of \cite{lampiris2018adding} comes with a cost -- extra transmitting antennas. However, we consider exactly the same setting as in the NMA scheme, i.e., single-antenna transmitters and receivers, and showed that the hypercube-based scheme can achieve the optimal sum-DoF and exponentially lower subpacketization levels without any extra cost. We will describe the difference of our hypercube-based design from \cite{lampiris2018adding} as follows.

In \cite{lampiris2018adding}, the main result is that in the cache-aided MISO broadcast channel with $L$ transmitter antennas and $\kr$ single-antenna receivers each equipped with a cache memory of size $M$ files, the sum DoF of $\kr\mu+L$ ($\mu \triangleq M/N$ is the normalized cache size) is achievable with subpacketization $F=\binom{\kr/L}{\kr\mu /L}$. The corresponding interference cancellation scheme therein first divides the receivers into $\kr/L$ groups each containing $L$ receivers and treat each group as a single receivers since they have the same cached contents. A direct translation of the shared-link coded caching scheme yields the a caching gain of $(\kr/L)\mu$. Then in each receiver group of $L$ receivers, a multi-antenna multiplexing gain of $L$ can be achieved using zero-forcing. As a result, each coded transmission can serve $\left( (\kr/L)\mu +1\right)L=\kr\mu +L$ different receivers interference-free, implying a sum-DoF of $\kr\mu +L$. Due to the cache placement property of the original coded caching scheme, a subpacketization level of $F=\binom{\kr/L}{\kr\mu/L}$ subfiles is inevitable. However, with multiple transmitting antennas, the inner-group interference among receivers can be effectively eliminated via zero-forcing. Due to the equivalence between the MISO broadcast channel with $L$ antennas to an interference channel with $L$ single-antenna transmitters, the proposed scheme can be readily carried out to the SISO channel with multiple transmitters.

In the case of $\kt \times \kr$ interference channel with transmitter  and receiver sides cache, measured by the caching parameter $\ttt=\kt\mu_T$ and $\ttr=\kr\mu_R$, the authors proposed to do the transmitter side cache placement in a circular way without extra splitting of the files. Since every $\ttt$ transmitters share some files, they can be treated as a MISO channel with $L=\ttt$ antennas and hence by applying the proposed scheme, a sum-DoF of $L+\kr\mu_R=\kt\mu_T + \kr\mu_R $ can be achieved while still preserving the relatively low subpacketization level. In summary, the subpacketization reduction comes from the effect of adding more transmitters as cost.
\fi


\section{Conclusion}

In this paper, we considered the cache-aided interference management problem where the transmitters and receivers are equipped with cache memories of certain sizes to pre-store parts of the contents. We adopt a new cache placement method called hypercube at both the transmitters' and receivers' sides. Based on the hypercube cache placement, we proposed a corresponding delivery scheme where the one-shot linear DoF of $\min\{\tron,\kr\}$ is achievable with exponentially less subpacketizations compared to the well-known NMA scheme. More specifically, via the design of the cache placement and the communication scheme, a set of $\tron$ packets can be delivered to the receivers simultaneously and interference-free, which is a joint effect of the zero-forcing (collaboration of transmitters via cache placement design at the transmitters' side) and cache cancellation (neutralization of known interference via the cache placement design at the receivers' side). The result shows that our proposed scheme can achieve exactly the same DoF performance as the NMA scheme while requiring significantly lower supacketization level. 


\appendices
\section{Proof of Lemma \ref{lemma_1}} 
\label{appendix_1}
First, we show that given a set of $|\cq|=Dt$ points (users) with $t$ dimensions and $D$ points in each dimension, the number of different hypercube permutations is  equal to 
$\left|\Pi^{\rm HCB}_{\mathcal{Q}}\right|=(D!)^t(t)!$
According to Definition \ref{defn_hcb_perm}, for a hypercube permutation $\mathbf{\pi}^{\rm HCB}$, the users belonging to the same dimension $\mathcal{U}_i$ can only appear in positions $p_{i,1},p_{i,2},\cdots,p_{i,D-1}$ such that $p_{i,j}\mod t=C_i,\,\forall j\in[0:D-1]$, where $C_i$ is a constant in terms of $j$ and $C_i\in[0:t-1]$. For two different dimensions $\mathcal{U}_{i_1}$ and $\mathcal{U}_{i_2}$, the corresponding modulo residues $C_{i_1}\neq C_{i_2}$ if $i_1\neq i_2$. As a result, $\{C_0,C_1,\cdots,C_{t-1}\}=\{0,1,\cdots,t-1\}$. Thus, given a group of users $\mathcal{U}_i$ and a prescribed modulo residue $C_i$, there are $D!$ ways to arrange these users to the corresponding set of positions $\left\{p_{i,j}:p_{i,j}\mod t=C_i,\,j\in [0:D-1]\right\}$. Since we have $t$ such user groups (dimensions), according to the multiplication principle, there are $(D!)^t$ ways to arrange all the users $\mathcal{Q}$ to the positions $\{p_{i,j}:p_{i,j}\mod t=C_i,\,j\in [0:D-1],\,i=0,1,\cdots,t-1\}$ under a prescribed modulo residue assignment. Since there are $t!$ different ways to assign the modulo residues $C_0,C_1,\cdots,C_{t-1}$ to the $t$ user groups, we conclude that $|\Pi^{\rm HCB}_{\mathcal{Q}}|=(D!)^{t}(t)!$. 

Now, for any $\pi\in \Pi^{\rm HCB}_{\mathcal{Q}}$, it is easy to see that there are $Dt-1$ other permutations in $\Pi^{\rm HCB}_{\mathcal{Q}}$ which are resulted from circularly shifting the elements of $\pi$. Since circular shifting is not allowed in the circular permutation, we have \begin{eqnarray}
\left\lvert\Pi^{\rm HCB,circ}_{\mathcal{Q}}\right\rvert=\frac{\left|\Pi^{\rm HCB}_{\mathcal{Q}}\right|}{Dt}=\frac{(D!)^{t}(t-1)!}{D}, 
\end{eqnarray} which completes the proof of Lemma \ref{lemma_1}.

\section{Proof of Lemma 2}
\label{sec: proof of lemma 2}
The proof of Lemma 2 can be completed by verifying the following two conditions: (1) For a specific receiver Rx$_j$, the number of packets it receives in the delivery phase equals the number of packets which are desired but have not been cached by Rx$_j$; (2) The number of packets received by all $K_R$ receivers equals the number of packets desired by them.
 
Each set in the union of (\ref{Eqn:grouping}) is composed of $t_T+t_R$ packets. The number of such sets is equal to
\begin{equation}
 D_T^{t_T}\binom{D_R}{\delta+1}^{t_R}\frac{\left((\delta+1)!)\right)^{t_R}(t_R-1)!}{\delta+1}.
\end{equation}
Therefore, the total number of packets in (\ref{Eqn:grouping}) is equal to
\begin{eqnarray}\label{1}
{ D_T^{t_T}\binom{D_R}{\delta+1}^{t_R}\frac{\left((\delta+1)!\right)^{t_R}(t_R-1)!}{\delta+1}(t_T+t_R)} 
=  { \frac{D_T^{t_T}K_R(D_R-1)!(D_R!)^{t_R-1}(t_R-1)!}{\left((D_R-\delta-1)!\right)^{t_R}}},
\end{eqnarray}
where we used the fact that $\delta =\frac{\ttt}{\ttr}$ and $\ttr=\frac{\kr}{\dr}$.

On the other hand, Rx$_j$, $j\in[K_R-1]$ has cached $D_T^{t_T}D_R^{t_R-1}$ subfiles in the prefetching phase, so the number of subfiles Rx$_j$ needs is equal to $D_T^{t_T}D_R^{t_R-1}(D_R-1)$. Since in the delivery phase, each desired subfile is further split into $\binom{D_R-2}{\delta-1}{\binom{D_R-1}{\delta}}^{t_R-1}\frac{(\delta!)^{t_R}}{\delta}(t_R-1)!$ packets, the total number of packets needed by Rx$_j$ is equal to 
\begin{eqnarray}\label{2}
&&{ D_T^{t_T}D_R^{t_R-1}(D_R-1)\binom{D_R-2}{\delta-1}{\binom{D_R-1}{\delta}}^{t_R-1}\frac{(\delta!)^{t_R}}{\delta}(t_R-1)!}\nonumber\\
&&={ \frac{D_T^{t_T}(D_R-1)!(D_R!)^{t_R-1}(t_R-1)!}{\left((D_R-\delta-1)!\right)^{t_R}}}.
\end{eqnarray}
Therefore, the total number of packets needed by all $K_R$ receivers is equal to
\begin{equation}\label{3}
 K_RD_T^{t_T}\frac{(D_R-1)!(D_R!)^{t_R-1}(t_R-1)!}{\left((D_R-\delta-1)!\right)^{t_R}},
\end{equation}  
which equals the total number of packets in (\ref{1}), implying that the set of packets needed by the receivers can be grouped into subsets of size $t_T+t_R$, verifying the second condition. Moreover, the number of packets received by Rx$_j$ in the delivery phase is equal to 
\begin{eqnarray}
{ D_T^{t_T}\binom{D_R-1}{\delta}{\binom{D_R}{\delta+1}}^{t_R-1}\frac{\left((\delta+1)!\right)^{t_R}(t_R-1)!}{\delta+1}} 
= {\frac{D_T^{t_T}(D_R-1)!(D_R!)^{t_R-1}(t_R-1)!}{\left((D_R-\delta-1)!\right)^{t_R}}}, 
\end{eqnarray}
which equals the number of packets calculated in (\ref{2}), verifying the first condition. As a result, the proof of Lemma 2 is complete.

\section{Proof of Theorem 2}
\label{sec: proof of Theorem 2}
We will first show that for any system parameters $K_T,K_R,M_T,M_R$ and $N$, which satisfy $K_T=D_Tt_T, K_R=D_Rt_R$ and $\delta=\frac{t_T}{t_R}\in \mathbb{Z}^+$, we have \emph{1)} $D_T^{t_T}< \binom{K_T}{t_T}$, \emph{2)} $D_R^{t_R}< \binom{K_R}{t_R}$, and \emph{3)} $\Delta_{\rm HCB}< \Delta_{\rm NMA}$. As a result, we obtain $G< 1$.

We first prove that $D_T^{t_T}< \binom{K_T}{t_T}$. For ease of notation, we denote $D_T$ as $d$ and $\ttt$ as $t$ for the time being. We have 
\begin{eqnarray}\label{step:1}
\frac{D_T^{t_T}}{\binom{K_T}{t_T}}=\frac{d^t}{\binom{dt}{t}}=\frac{d^tt!}{dt(dt-1)(dt-2)\cdots\left(dt-(t-1)\right)} 
= \left(\frac{t}{t}\right)\left(\frac{t-1}{t-\frac{1}{d}}\right)\cdots\left(\frac{t-(t-1)}{t-\frac{t-1}{d}}\right).
\end{eqnarray}
Since we have assumed that $d\geq \delta+1\geq2$ where $\delta\geq 1$, it can be seen that $t-i\leq t-{i}/{d},\forall i\in[t-1]$, implying that each individual term in the RHS of  (\ref{step:1}) is less than 1. As a result, the product is less than 1, implying $D_T^{t_T}< \binom{K_T}{t_T}$. Similarly, we can prove $D_R^{t_R}< \binom{K_R}{t_R}$.

Next we prove $\Delta_{\rm HCB}< \Delta_{\rm NMA}$. Denote $t_R$ as $t$ and $D_R$ as $d$, we have $t_T=\delta t_R=\delta t$. Thus, $\Delta_{\rm HCB}$ and $\Delta_{\rm NMA}$ can be simplified as   
\begin{eqnarray}
\Delta_{\rm HCB} &=&\binom{d-2}{\delta-1}\binom{d-1}{\delta}^{t-1}\frac{(\delta !)^t}{\delta}(t-1)! 
= \frac{\left((d-1)!\right)^t(t-1)!}{\left((d-\delta-1)!\right)^t(d-1)},\\
\Delta_{\rm NMA} &=&\binom{dt-t-1}{\delta t-1}(\delta t-1)!t!=\frac{(dt-t-1)!t!}{\left((d-\delta-1)t\right)!}.
\end{eqnarray}
Therefore,
\begin{eqnarray}
\frac{\Delta_{\rm NMA}}{\Delta_{\rm HCB}}&=&\frac{\left((d-\delta -1)!\right)^t\left((d-1)t\right)!}{\left((d-\delta -1)t\right)!\left((d-1)!\right)^t} 
= \frac{\prod_{i=0}^{\delta t-1}\left((d-1)t-i\right)}{\left(\prod_{i=0}^{\delta-1}(d-1-i)\right)^t} 
= \lambda_0\lambda_1\cdots\lambda_{t-1}, 
\end{eqnarray}
in which the parameter $\lambda_k$ is defined as 
\begin{eqnarray}\label{lambda_k}
\lambda_k\triangleq\frac{\prod_{i=k\delta}^{(k+1)\delta -1}\left((d-1)t-i\right)}{\prod_{i=0}^{\delta-1}(d-1-i)},\quad \forall k\in[t-1].
\end{eqnarray} 
Note that $\lambda_0>\lambda_1>\cdots>\lambda_{t-1}$. Next we show that $\lambda_{t-1}\geq 1$. From (\ref{lambda_k}), we have  
\begin{eqnarray}\label{Lambda_0}
\lambda_{t-1}&=&\frac{\prod_{i=(t-1)\delta}^{\delta t -1}\left((d-1)t-i\right)}{\prod_{i=0}^{\delta-1}(d-1-i)} 
= \prod_{i=0}^{\delta-1}\left(t-\frac{(\delta-i)(t-1)}{d-1-i}\right) \nonumber\\
&\overset{({\rm a})}{\geq}&\prod_{i=0}^{\delta-1}\left(t-\frac{(\delta-i)(t-1)}{\delta+1-1-i}\right) 
=\prod_{i=0}^{\delta-1}\left(t-(t-1)\right) 
=1,
\end{eqnarray} 
where in (a) we used the assumption that $d\geq \delta+1$. Hence, we obtain that $\lambda_{t-1}\geq1$. Since $\lambda_0>\lambda_1>\cdots>\lambda_{t-1}\geq 1$, we have $\frac{\Delta_{\rm NMA}}{\Delta_{\rm HCB}}=\lambda_0\lambda_1\cdots\lambda_{t-1}> 1$, implying $\Delta_{\rm HCB}< \Delta_{\rm NMA}$. Combining the above results, we conclude that the multiplicative gap $G$ is strictly less than 1 for any system parameters, i.e.,
\be
G=\frac{\dt^{\ttt}\dr^{\ttr}}{\binom{\kt}{\ttt}\binom{\kr}{\ttr}}\cdot\frac{\Delta_{\rm HCB}}{\Delta_{\rm NMA}}<1.
\ee
This proof also indicates that the hypercube based scheme requires less number of subfiles per file in the prefetching phase and and less number of packets per subfile in the delivery phase than the NMA scheme. 


{Next we prove the upper bound on $G(d,t,\delta)\le  C_0\left(\frac{C_1}{t} \right)^t\frac{1}   {t^{(\delta -1) t-1}},  $ in Theorem \ref{theorem:2}. We set $\dt=\dr =d$ and $\ttr =t,\ttt=\delta\ttr=\delta t$. Since the normalized per-Tx/Rx cache memory $N/M_T=NM_R=1/d$ is fixed, we see that increasing the caching parameter $t=K_RM_R/N=K_R/d$ is equivalent to increasing the number of transmitters/receivers (note that $K_T=\delta K_R$). The key technique we used to derive the upper bound is \emph{Stirling's approximation} which states that for any $ n\in\mathbb{Z}^+$, $n!$ is bounded by  
\begin{eqnarray}
\label{eq: stirling approx}
\sqrt{2\pi } n^{n+\frac{1}{2}}e^{-n} \le n! \le e n^{n+\frac{1}{2}}e^{-n}
\end{eqnarray}

With the above setting, we have
\begin{subequations}
\begin{align}
G(d,t,\delta)
& =\frac{ D_T^{t_T}D_R^{t_R} \binom{D_R-2}{\delta-1}{\binom{D_R-1}{\delta}}^{t_R-1}\frac{(\delta!)^{t_R}}{\delta}(t_R-1)!   }{\binom{K_T}{t_T}\binom{K_R}{t_R}\binom{K_R-t_R-1}{t_T-1}(t_T-1)!t_R!  }\\
&=\frac{d^{\delta t}d^{t} \binom{d-2}{\delta-1}\binom{d-1}{\delta}^{t-1}(\delta !)^{t-1}(\delta -1)!(t-1)!      }{ \binom{\delta dt}{\delta t}\binom{dt}{t}\binom{(d-1)t-1}{\delta t-1} (\delta t -1)!t!        }\label{eq: step 1}\\
& = \frac{ d^{(\delta +1)t} \frac{(d-2)!}{(d-\delta -1)!} \left(  \frac{(d-1)!}{(d-\delta -1)!}     \right)^{t-1}(t-1)!  }{ \frac{(\delta dt)!}{(\delta t)![(d-1)\delta t]!}\frac{(dt)!   }{[(d-1)t]!} \frac{[(d-1)t-1]!}{[(d-\delta -1)t]!    }      }\label{eq: step 2}
\end{align}
\end{subequations}
in which  (\ref{eq: step 1}) is due to $(\delta !)^{t}/\delta = (\delta! )^{t-1}(\delta -1)$. We now aim to obtain an upper bound on $G(d,t,\delta)$ by finding an upper bound on the numerator of  (\ref{eq: step 2}) and a lower bound to the denominator using Stirling's bounds on the factorials.

For the numerator, we can find an upper bound $(t-1)!\le (t-1)^{t-{1}/{2}}e^{2-t}$ using (\ref{eq: stirling approx}). The denominator of (\ref{eq: step 2}) consists of three terms each of which can be lower bounded and therefore the denominator can be lower bounded. In particular, we show how to lower bound the first term  
$\frac{(\delta dt)!}{(\delta t)![(d-1)\delta t]!}$. The idea is to lower bound the numerator and upper bound the denominator respectively. More specifically, using (\ref{eq: stirling approx}), we obtain  
\begin{eqnarray}
 (\delta dt)!  &\ge &\sqrt{2\pi} (\delta dt)^{\delta dt +{1}/{2}}   ,\\
 (\delta t)! &\le & (\delta t)^{\delta t+1/2}e^{1-\delta t},\\
\left((d-1)\delta t\right)! & \le &  \left(d-1)\delta t\right)^{(d-1)\delta  t+1/2}e^{1-(d-1)\delta t},
\end{eqnarray}
which together implies 
\begin{equation}
\frac{(\delta dt)!}{(\delta t)![(d-1)\delta t]!} \ge 
\sqrt{2\pi \delta}\left( \frac{d}{d-1}  \right)^{(d-1)\delta t +1/2}d^{\delta t}t^{-1/2}.
\end{equation}
Similarly, the second and third terms of the denominator of (\ref{eq: step 2}) can be lower bounded by
\begin{eqnarray}
\frac{(dt)!}{[(d-1)t]!}  &\ge & \sqrt{2\pi}\left(   \frac{d}{d-1}\right)^{(d-1)t+1/2} d^t t^t e^{-(t+1)} ,\\
\frac{[(d-1)t-1]!}{[(d-\delta -1)t]!} &\ge & \sqrt{2\pi} \left(\frac{d-1}{d-\delta -1} \right)^{(d-\delta -1)t+1/2}(d-1)^{\delta t -1} t^{\delta t-1}e^{-(\delta t+1)}.
\end{eqnarray}
Combining the above bounds and continuing with (\ref{eq: step 2}), we obtain an upper bound as
\begin{subequations}
\begin{align}
G(d,t,\delta) &\le   \frac{  C_0d^{(\delta +1)t} \left(  \frac{(d-1)!}{(d-\delta -1)!}     \right)^{t-1}(t-1)^{t-1/2}e^{2-t}  }{
 \left(  \frac{d}{d-1} \right)^{(d-1)(\delta+1)t}  \left(  \frac{d-1}{d-\delta -1}  \right)^{(d-\delta -1)t}  d^{(\delta +1 )t} (d-1)^{\delta t}t^{(\delta +1)t-3/2}e^{-(\delta +1)t} }\\
& \le        \frac{{C_0}{C_1}^{-1}e^2  C_1^{t}e^{\delta t}  }{
 C_2^t C_3^t (d-1)^{\delta t}t^{\delta t-1} }\label{eq: step 3}\\
&  =       \frac{{C_0}{C_1}^{-1}e^2  C_1^{t}e^{\delta t}  }{
 C_2^t C_3^t   (d-1)^{\delta t}t^{\delta t-1} }\\
&  = \frac{C_0e^2}{C_1}\left(\frac{C_1e^{\delta}}{C_2C_3(d-1)^{\delta}}  \right)^t\frac{1}{t^{\delta t-1}}\\
 & = \frac{C_0e^2}{C_1}\left(\frac{C_4}{t}  \right)^t\frac{1}{t^{(\delta-1) t-1}}\\
 &\overset{t\ge C_4}{\le }\frac{C_0e^2}{C_1} \frac{1}{t^{(\delta-1) t-1}}\label{eq: step 4}
\end{align}
\end{subequations}
in which the parameters (not depending on $t$) are $C_0=\frac{(d-2)!}{(d-\delta -1)!} /\left(\sqrt{\frac{d-1}{\delta}}\left( \frac{2\pi}{d-1} \right)^{3/2}e^{-4}\right), C_1 =\frac{(d-1)!}{(d-\delta -1)!}, C_2 = \left(  \frac{d}{d-1} \right)^{(d-1)(\delta+1)}, C_3=\left(  \frac{d-1}{d-\delta -1}  \right)^{(d-\delta -1)}$ and $C_4 =\frac{C_1e^{\delta}}{C_2C_3(d-1)^{\delta}}$. (\ref{eq: step 3}) is due to $(t-1)^{t-1/2}<t^{t-1/2}$. Sine we are considering the asymptotic scaling of $G(d,t,\delta)$, (\ref{eq: step 4}) holds as long as $t\ge C_4$. Moreover, for fixed $d$ and $\delta$, we will have the scaling $G(d,t,\delta)=O(1/t^{(\delta -1)t-1})$ as $t\to \infty$. As a result, the proof of Theorem \ref{theorem:2} is complete.}

\bibliographystyle{IEEEbib}
\bibliography{references_d2d}

\ifCLASSOPTIONcaptionsoff
  \newpage
\fi

\end{document}